\documentclass[%
 reprint,
 amsmath,amssymb,
 aps,
prb,
]{revtex4-2}
\pdfoutput=1
\usepackage{multirow}
\usepackage{graphicx}
\usepackage{dcolumn}
\usepackage{bm}
\usepackage{hyperref}

\begin{document}

\preprint{APS/123-QED}

\title{Description of two-dimensional altermagnetism: Categorization using spin group theory}

\author{Sike Zeng}
\author{Yu-Jun Zhao}%
\email{zhaoyj@scut.edu.cn}
\affiliation{%
 Department of Physics, South China University of Technology, Guangzhou 510640, China
}%


\date{\today}

\begin{abstract}
Altermagnetism, recently spotlighted in condensed matter physics, presents captivating physical properties and holds promise for spintronics applications. This study delves into the theoretical description and categorization of two-dimensional altermagnetism using spin group theory. Employing spin-group formalism, we establish seven distinct spin layer groups, extending beyond the conventional five spin Laue groups, to describe two-dimensional altermagnetism. Utilizing these findings, we classify previously reported two-dimensional altermagnets and identify novel materials exhibiting altermagnetism. Specifically, monolayer MnTeMoO$_6$ and VP$_2$H$_8$(NO$_4$)$_2$ are predicted to be two-dimensional altermagnets. Furthermore, we scrutinize their spin-momentum locking characteristics through symmetry analysis and density functional theory calculations, substantiating their altermagnetic properties.

\end{abstract}

\maketitle


\section{INTRODUCTION}
Altermagnetism, characterized by collinear-compensated magnetic order in real space and time-reversal symmetry breaking in reciprocal space, has recently attracted considerable attention in condensed matter physics\cite{vsmejkal2022emerging,zhou2024crystal,mazin2024altermagnetism,chen2023spin}, for its intriguing physical properties and promising application in spintronics. This phenomenon was proposed in scientific literature as early as  2019\cite{ma2021multifunctional,mazin2021prediction,hayami2019momentum,vsmejkal2020crystal,PhysRevB.102.014422}, and subsequently formalized with the name `altermagnetism' in 2022\cite{vsmejkal2022beyond}. In the realm of nonrelativistic spin groups\cite{brinkman1966theory,litvin1974spin,litvin1977spin}, altermagnetism has emerged as a distinct third magnetic state, characterized by the connection between opposing sublattices through rotational or mirror symmetries, rather than through translational or inversional symmetries, leading to the disruption of \textbf{\emph{PT}} symmetry\cite{vsmejkal2022beyond}. Due to its unique spin-momentum locked electronic structure, there are a number of unconventional anomalous magnetic response predicted in altermagnets, such as anomalous Hall effect\cite{vsmejkal2022anomalous,vsmejkal2020crystal} and Kerr effect\cite{samanta2020crystal,zhou2021crystal}, which have been verified by experiments\cite{bai2022observation}. Moreover, very recently, the spin-splitting electronic structure has been reported experimentally in MnTe, through angle-resolved photoemission spectroscopy measures, which provides direct evidence for altermagnetism\cite{lee2024broken}. In terms of application, due to the advantage of no stray field, THz spin dynamics and strong T-symmetry breaking, altermagnet hold great promise for applications in spintronics. It has been suggested that giant/tunneling magnetoresistance (GMR/TMR) effect\cite{PhysRevX.12.011028} and spin-splitter torque\cite{PhysRevLett.126.127701} can be generated in altermagnets.

So far, a wide range of materials has been classified as altermagnets, such as RuO$_2$\cite{vsmejkal2020crystal}, MnTe\cite{lee2024broken,PhysRevLett.130.036702}, MnF$_2$\cite{PhysRevB.102.014422}, and so on. However, the majority of reported altermagnets are three-dimensional (3D), while there are few reports on two-dimensional (2D) altermagnets. It has been shown that monolayer MnP(S,Se)$_3$ can transform from antiferromagnetism to altermagnetism by applying an electric field or through a Janus structure, which breaks the inversion symmetry between one sublattice and another\cite{mazin2023induced}. Furthermore, some 2D altermagnets have been theoretically predicted, such as V$_2$Te$_2$O\cite{PhysRevB.108.024410} and RuF$_4$\cite{sodequist2024two}. It has been proposed that magnon-mediated superconductivity may occur in 2D altermagnets and the critical temperature can be enhanced by tuning the chemical potential\cite{PhysRevB.108.22442}. However, to date, no specific material exhibiting altermagnetism have been demonstrated to show this result experimentally. Therefore, the search for 2D materials with altermagnetism is crucial for a deeper understanding of the  fundamental physical properties inherent in 2D altermagnets.

On the other hand, 3D altermagnetism has been described by spin Laue group in the previous research\cite{vsmejkal2022beyond}, while how to describe 2D altermagnetism has not been solved. On the basis of Laue group and spin-group formalism, which considers symmetry transformation in decoupled real and spin space, it is suggested that there are 32 nontrivial spin Laue groups, which correspond to three distinct magnetic phases. According to the characteristics of spin-momentum locking, the ten nontrivial spin Laue groups of altermagnetism are classified into two types, the plane and bulk. Five plane spin Laue groups are considered to appear in both two-dimensional and three-dimensional crystals, whereas the five bulk spin Laue groups are thought to appear only in three-dimensional crystals. Here, we discover that there are seven spin groups to describe the spin-momentum locking in 2D altermagnet, rather than five spin Laue groups. The unexpected spin groups were previously classified as bulk type.

In fact, the significance of altermagnetism in 2D materials remains an open question. As stated by the Mermin-Wagner theorem\cite{PhysRevLett.17.1133}, magnetic anisotropy originating from spin-orbit coupling (SOC) is essential for magnetic order in 2D materials at finite temperatures. The impact of SOC on RuF$_4$, identified as a 2D altermagnet, has been discussed. It has been proposed that in the presence of SOC, 2D altermagnets may exhibit weak ferromagnetism and SOC-induced band splitting\cite{milivojevic2024interplay}. Nevertheless, in our theoretical analysis and DFT calculations, we have neglected the SOC effect, given the nonrelativistic origin of altermagnetism. However, to be more rigorous, we also provide a comparison of band structures with and without SOC in Appendix B. 

In this work, we focus on the description and classification of 2D altermagnetism. Utilizing a method similar to that used to deduce the spin group theory, we have explored the two-dimensional spin layer group, which is grounded in the layer group concept. The spin layer groups are clearly beneficial for the search for and further understanding of 2D altermagnetism. We opt for layer groups, rather than point groups, as the symmetry transformations in real space for two main reasons. First, the 2D materials we commonly study are not two-dimensional strictly but quasi two-dimensional. Secondly, employing point groups can introduce ambiguity. Consequently, we believe that layer groups provide a more accurate symmetry description for 2D materials. To verify our results, we employ spin layer groups that describe altermagnetism to search for potential materials with 2D altermagntism in \emph{material project} database\cite{10.1063/1.4812323} and to classify the reported 2D altermagnets. 

The paper is organized as follows: in the second section, we illustrate the process of deducing spin layer groups. In the third section, we provide some examples of 2D altermagnets to verify our result and analyze their spin-momentum locking properties. Finally, we present our conclusions and offer some prospects about 2D altermagnetism.

\section{DERIVATION OF SPIN LAYER GROUP}

\begin{figure}
	
	\centering
	\includegraphics[width=\linewidth]{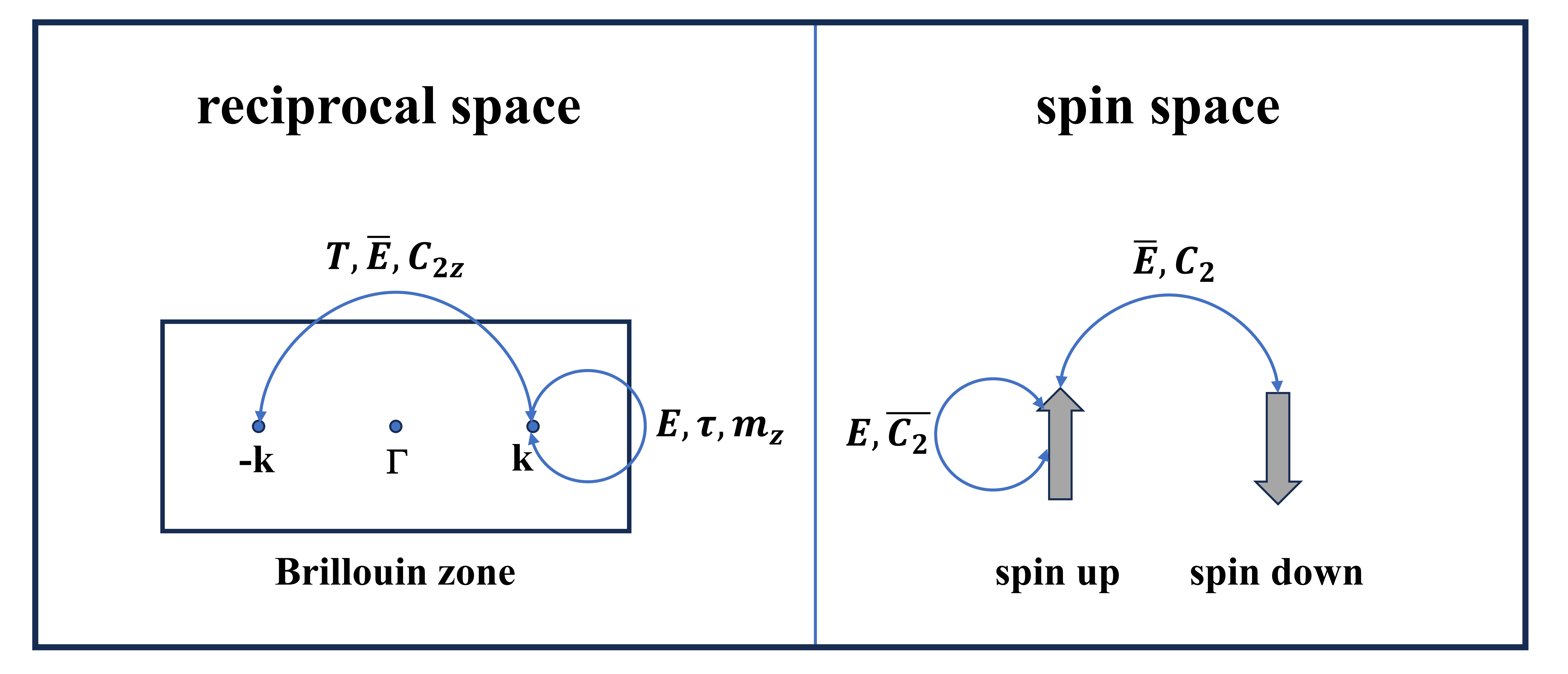}
	\caption{\label{fig:figure0}The schematic diagram of the effect of various spin symmetries on the $\varepsilon(s,\textbf{k})$ spectrum. In reciprocal space, \emph{T}, $\overline{E}$ and $C_{2z}$ represent time-inversion symmetry, space-inversion symmetry and 2-fold rotational symmetry around \emph{z}-axis  respectively. They will map k-vector to its opposite number. $E$, $\tau$ and $m_z$ represent identity, translation symmetry and mirror symmetry parallel to \emph{xy}-plane, and they will map k-vector to itself. In spin space, $\overline{E}$ and $C_{2}$ represent inversion symmetry and 2-fold rotational symmetry an axis perpendicular to the spins respectively. They will transform the spin to the opposite one, while identity symmetry $E$ and a 2-fold rotation around an axis perpendicular to the spins, combined with inversion symmetry of spin space, $\overline{C_2}$ will make it invariant.}
	
\end{figure}

As is known to all, spin group is expressed as the direct product $r_s\otimes R_s$, where $r_s$ represents spin-only group containing symmetry transformations acting on spin space, and $R_s$ represents nontrivial spin group containing pairs of transformations $[R_i||R_j]$, in which the transformations on the left of the double bar only act on the spin space and those on the right of the double bar only act on the real space. For collinear spin arrangements, the spin-only group mainly include two transformations\cite{vsmejkal2022beyond}. One is all rotations in spin space around the common axis of spins $C_\infty$, which make spin a good quantum number, so the band structure can be strictly separated into spin-up and spin-down channels. The other is a 2-fold rotation around an axis perpendicular to the spins, combined with inversion symmetry of spin space, $\overline{C_2}$, which is always accompanied by time-inversion symmetry(\emph{T}), and can be written as $[\overline{C_2}||\emph{T}]$. When we apply it on the spin and momentum-dependent bands, $[\overline{C_2}||\emph{T}]\varepsilon(s,\textbf{k})=\varepsilon(s,-\textbf{k})$ can be obtained. We provide a schematic diagram in FIG.\ref{fig:figure0}, for a better understanding of the effect of various spin symmetries on the $\varepsilon(s,\textbf{k})$ spectrum. Therefore, this transformation is equivalent to the real-space inversion in reciprocal space. Meanwhile, because this transformation makes the whole system invariant, for all collinear magnets, we can obtain $[\overline{C_2}||\emph{T}]\varepsilon(s,\textbf{k})=\varepsilon(s,\textbf{k})$, and thus $\varepsilon(s,\textbf{k})=\varepsilon(s,-\textbf{k})$. It can thus be concluded that under real-space inversion, the nonrelativistic band structure remains invariant for all collinear magnets, regardless of whether the system possesses real-space inversion symmetry. This conclusion is significant, as you can see later in this article, greatly simplifies the derivation of the spin layer group.

\begin{table*}
	\caption{\label{tab:table1}
		All reciprocal-space symmetry corresponding to layer groups. The plus (+) and minus (-)  denote the counterclockwise and clockwise rotations, respectively. The bar above the symbol denotes inversion symmetry. For example, $\overline{C_{4z}^+}$ denotes 90 degree rotation counterclockwise with space inversion. A sequence of three numbers denotes the direction of axis or plane. The direct product is denoted by $\otimes$.
	}
	\begin{ruledtabular}
		\begin{tabular}{ccc}
			\textrm{reciprocal-space group}&
			\textrm{Layer group}&
			\textrm{reciprocal-space symmetry}\\
			\hline
			1 & 1-2 & $E ,\overline{E}$ \\
			2 & 3-7 & $E,\overline{E} ,C_{2z},m_z$\\
			3 & 8-18 & $E,\overline{E} ,C_{2x},m_x$\\
			4 & 19-48 & $E,\overline{E} ,C_{2z},m_z, C_{2x}, m_x, C_{2y}, m_y$\\
			5 & 49-52 & $E,\overline{E} ,C_{2z},m_z, C_{4z}^+, C_{4z}^-, \overline{C_{4z}^+}, \overline{C_{4z}^-}$\\
			6 & 53-64 & $E,\overline{E} ,C_{2z},m_z, C_{4z}^+, C_{4z}^-, \overline{C_{4z}^+}, \overline{C_{4z}^-}, C_{2x}, m_x, C_{2y}, m_y, C_2^{110}, m_{110}, C_2^{1\overline{1}0}, m_{1\overline{1}0}$\\
			7 & 65-66 & $E,\overline{E}, C_{3z}^+, C_{3z}^-, \overline{C_{3z}^+}, \overline{C_{3z}^-}$\\
			\multirow{2}*8 & \multirow{2}*{67-72} & $E,\overline{E}, C_{3z}^+, C_{3z}^-, \overline{C_{3z}^+}, \overline{C_{3z}^-},  C_2^{110}, m_{110},C_{2x}, C_{2y}, m_{\overline{2}10}, m_{\overline{1}20}$\\
			 & & $(E,\overline{E}, C_{3z}^+, C_{3z}^-, \overline{C_{3z}^+}, \overline{C_{3z}^-}, C_2^{1\overline{1}0}, C_2^{120}, C_2^{210}, m_{1\overline{1}0}, m_x, m_y)$\\
			9 & 73-75 &  $E,\overline{E}, C_{3z}^+, C_{3z}^-, \overline{C_{3z}^+}, \overline{C_{3z}^-}, C_{2z},m_z,C_{6z}^+,\overline{C_{6z}^+},C_{6z}^-,\overline{C_{6z}^-} $\\
			10 & 76-80 &  $\{E,\overline{E}\}\otimes\{E, C_{3z}^+, C_{3z}^-, C_{2z},C_{6z}^+, C_{6z}^-,C_2^{110},C_{2x}, C_{2y},  C_2^{1\overline{1}0}, C_2^{120}, C_2^{210}\} $
		\end{tabular}
	\end{ruledtabular}
\end{table*}

We now focus on the nontrivial spin group. The nontrivial spin group involves pairs of transformations $[R_i||R_j]$, which act independently on spin space and real space. Regarding the spin-space symmetry, for collinear spin arrangements, we select two transformations. One is the identity \emph{E} of spin space, and the other is a 2-fold rotation around an axis perpendicular to the spins, denoted as $C_2$. Alternatively, spin-space inversion can also be chosen. These options result in the formation of two spin-space groups, $S_1=\{E\}$ and  $S_2=\{E,C_2\}$. 

Next, we address the real-space transformations within the nontrivial spin group. First of all, we assume that the 2D material under consideration lies parallel to the \emph{xy}-plane. For the reasons mentioned earlier, we select the layer group as the real-space group to construct nontrivial spin layer group. Detailed information on layer groups can be found in a referenced book\cite{kopsky2002international}. As the prominent property of altermagnetism we focus on is spin-momentum locking, which is independent of real-space translation symmetry, we only consider the point-group symmetry operator of layer-group symmetry. For example, the glide reflection through the \emph{xy}-plane is regarded as mirror symmetry through the \emph{xy}-plane regardless of the translation direction. Similarly, the screw rotation around \emph{x}-axis is regarded as a 2-fold rotation around \emph{x}-axis. As stated above, for all collinear magnets with or without real-space inversion symmetry, the nonrelativistic band structure will be invariant, when applied with real-space inversion symmetry operations. To avoid confusing between layer-group symmetry and the symmetry that keeps band structure invariant, we refer to the latter as reciprocal-space symmetry. Through the direct product of the space-inversion group and the layer groups, we can obtain all reciprocal-space groups that correspond to the eighty layer groups, as listed in Table \ref{tab:table1}. In fact, these groups belong to the Laue groups.

According to the isomorphism theorem\cite{litvin1974spin}, which implies the decomposition with the same number of cosets for the two groups, the nontrivial spin layer group can be constructed into three types, corresponding to three magnetic phase respectively. Selecting \( S_1 \) as the transformation for the spin space, there exists a unique scenario in which the nontrivial spin layer group \( R_1 = [S_1 || G] \) can be derived, with \( G \) representing the layer groups. They describe nonzero magnetization and band structure with broken time-inversion symmetry, corresponding to ferromagnetism. 

\begin{table*}
	\renewcommand{\arraystretch}{3.8}
	\caption{\label{tab:table2}All nontrivial spin layer groups, corresponding layer groups, corresponding reciprocal-space group(RG), the basic characteristic of spin-momentum locking and material candidates. Here we adopt Litvin's notation of the spin groups. The sign * denotes that the material is predicted by symmetry analysis in this work and the evidence of altermagnetism can be obtained in Section \ref{sec: level 3}. The other materials have been reported in the previous research.}
	\begin{ruledtabular}
		\begin{tabular}{ccccc}
			\textrm{Nontrivial spin layer group}&
			\textrm{Spin-momentum locking $(k_x,k_y)$}&
			\textrm{Layer group}&
			\textrm{RG}&
			\textrm{Material candidate}\\ 
			\hline
			$^{2}2/^{2}m_x$ & \multirow{2}*{\includegraphics[width=3cm]{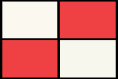}} & 8-18 & 3 & RuF$_4$\cite{sodequist2024two,milivojevic2024interplay}\\ 
			$^{2}m^{2}m^{1}m$ & & 19-48& 4 & MnTeMoO$_6^*$\\ \hline
			$^{2}4/^{1}m$ & \multirow{2}*{\includegraphics[width=2.5cm]{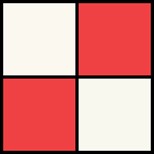}} & 49-52 & 5& \\
			$^{2}4/^{1}m^{2}m^{1}m$ & & 53-64 & 6 & V$_2$Se$_2$O\cite{ma2021multifunctional}, V$_2$Te$_2$O\cite{PhysRevB.108.024410}, Cr$_2$O$_2$\cite{chen2023giant,guo2023quantum}\\ \hline
			\multirow{2}*{$^{1}4/^{1}m^{2}m^{2}m$} & \multirow{2}*{\includegraphics[width=2.5cm]{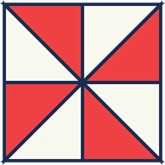}} &
			\multirow{2}*{53-64} & \multirow{2}*6 &
			\multirow{2}*{VP$_2$H$_8$(NO$_4$)$_2^*$}  \\
			& & & & \\ \hline
			$^{1}\overline{3}$$^{2}m$ &  \multirow{2}*{\includegraphics[width=3cm]{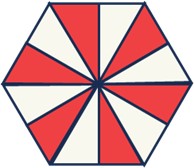}} & 67-72 & 8 & Mn$_2$P$_2$S$_3$Se$_3$\cite{mazin2023induced} \\
			$^{1}6/$$^{1}m$$^{2}m$$^{2}m$ & & 76-80& 10 & \\
		\end{tabular}
	\end{ruledtabular}
\end{table*}

Similarly, if we choose $S_2$ as the spin-space transformation, two cases will appear. One is that the nontrivial spin layer group is given by $R_2=[S_2||G]=[E||G]+[C_2||G]$, which describes zero magnetization and spin-degenerate band structure with time-inversion symmetry, corresponding to antiferromagnetism. The other is that the nontrivial spin layer group is given by $R_3=[E||H]+[C_2 ||G-H]$, where H is a halving subgroup of layer group G. They describe zero magnetization and band structure with broken time-inversion symmetry, corresponding to altermagnetism, which we will focus on later. The transformation associated with spin-space identity will link atoms within the same sublattice, whereas the transformation involving a 2-fold rotation in spin space will connect atoms from one sublattice to those of the opposite sublattice.

In 2D system, certain symmetries protect the spin degeneracy of the nonrelativistic band structure for all \textbf{k}-vector in the whole Brillouin zone. First, $[C_2||\tau]$ symmetry obviously implies spin degeneracy of band structure, where $\tau$ is translation symmetry connecting atoms of the opposite sublattices. As the spin degeneracy is independent of real-space translation symmetry, $[C_2||\tau]$ is equivalent to $[C_2||E]$, which means $[C_2||E]\varepsilon(s,\textbf{k})=\varepsilon(-s,\textbf{k})=\varepsilon(s,\textbf{k})$. Moreover, the spin degeneracy is also protected by the inversion symmetry connecting atoms of the opposite sublattices, i.e. $[C_2||\overline{E}]$. This is because $[\overline{C_2}||T]$ is one of spin-only symmetry and the spin (layer) group is a direct product of spin-only group and nontrivial spin (layer) group, i.e. $[C_2||\overline{E}][\overline{C_2}||T]=[\overline{E}||T\overline{E}]$, which is \textbf{\emph{PT}} symmetry to protect Kramers spin degeneracy. These results have been reported previously\cite{vsmejkal2022beyond}. For 2D symmetry, however, there are two extra symmetries to protect Kramers spin degeneracy. The first one is $[C_2||m_z]$, which represents that the atoms of the opposite sublattices can be connected by the mirror symmetry through the  \emph{xy}-plane. In this case, we can obtain $[C_2||m_z]\varepsilon(s,\textbf{k})=\varepsilon(-s,\textbf{k})$. Meanwhile, $[C_2||m_z]$ is also the symmetry of system, i.e., $[C_2||m_z]\varepsilon(s,\textbf{k})=\varepsilon(s,\textbf{k})$. Therefore, spin degeneracy of the band structure will appear in the materials with symmetry $[C_2||m_z]$. The second one is $[C_2||C_{2z}]$, which represents the atoms of the opposite sublattices can be connected by a 2-fold rotation around \emph{z}-axis. It can be found that the product of $[C_2||C_{2z}]$ and spin-only symmetry $[\overline{C_2}||T]$ equals to $[\overline{E}||TC_{2z}]$. This means that  $[\overline{E}||TC_{2z}]\varepsilon(s,\textbf{k})=\varepsilon(-s,\textbf{k})=\varepsilon(s,\textbf{k})$, so spin-degeneracy band structure is protected in the 2D materials with symmetry $[\overline{E}||TC_{2z}]$. These symmetries have also been suggested to protect spin-degeneracy in recent work\cite{sodequist2024two}.These four symmetries must be excluded, when we deduce nontrivial spin layer group for altermagnetism. If a 2D material possesses any of the crystal symmetries $\{\tau,m_z,\overline{E},C_{2z}\}$ in real space, this symmetry must be accompanied by spin-space identity in the altermagnetic phase. Therefore, the atoms of the opposite sublattices can not be connected by any of the crystal symmetries $\{\tau,m_z,\overline{E},C_{2z}\}$ in 2D altermagnets.

Then, we are going to focus on constructing the nontrivial spin layer groups to describe altermagnetism, starting from the formalism $R_3=[E||H]+[C_2||G-H]$ and the reciprocal-space groups coming from layer groups. We finally obtain seven nontrivial spin layer groups, as listed in Table \ref{tab:table2}. Here we adopt Litvin's notation of the spin groups\cite{litvin1977spin}, in which upper index 1 represents the identity symmetry in spin space and the upper index 2 represents the rotation symmetry $C_2$ in spin space. Notably, some reciprocal-space groups are incapable of forming nontrivial spin layer group for altermagnetism, as they lack the appropriate halving subgroup. This deficiency necessitates that the symmetries, $E,m_z,\overline{E}$ and $C_{2z}$, should be accompanied by spin-space identity to prevent spin degeneracy if these symmetries are present in the real-space group. Therefore, not all layer groups can be constructed to nontrivial spin layer group, as listed in Table \ref{tab:table2}. For example, the second reciprocal-space group in Table \ref{tab:table1} possesses four symmetries $\{E,m_z,\overline{E},C_{2z}\}$, all of which must be accompanied by spin-space identity when constructing nontrivial spin layer group for altermagnetism. Therefore, this reciprocal-space group can not be constructed to nontrivial spin layer group. It is worth noting that the five spin Laue groups describing 2D materials in previous research are all included within the seven spin layer groups. Interestingly, two results not previously categorized as plane-type altermagnetism have emerged. One such case is the nontrivial spin layer group $^{2}2/^{2}m_x$, i.e., $R_3=[E||\{E,\overline{E}\}]+[C_2||\{C_{2x},m_x\}]$, which exhibits the same spin-momentum locking characteristics as the group $^{2}m^{2}m^{1}m$, previously identified as plane-type altermagnetism. We also find that it can be understood as the spin Laue group $^{2}2/^{2}m$ in the case where the 2-fold rotation symmetry is around \emph{x}-axis. This is because the symmetry $[C_2||m_z]$, which preserve spin-degeneracy band in the whole Brillouin zone, disappears in this case. By symmetry analysis, we find that the material RuF$_4$ exhibits the symmetries in this group, as described in Section \ref{sec: level 3}. The other is group $^{1}\overline{3}$$^{2}m$, i.e., 
$R_3=[E||\{E,\overline{E}, C_{3z}^+, C_{3z}^-, \overline{C_{3z}^+}, \overline{C_{3z}^-}\}]
+[C_2||\{C_2^{110}, m_{110},C_{2x}, C_{2y}, m_{\overline{2}10}, m_{\overline{1}20}\}] $, 
which has been classified as bulk-type altermagnetism in three-dimensional altermagnetism. However, it is the absence of the symmetry $[C_2||m_z]$ and $[C_2||C_{2z}]$ that make it appear in the seven spin layer groups. In two-dimensional altermagnetism, it share the same spin-momentum locking properties as the group $^{1}6 /^{1}m$$^{2}m$$^{2}m$, given that the symmetry $\overline{C_{3z}^+}$ is equivalent to $C_{6z}^-$ in two-dimensional reciprocal space. Additionally, we have identified that  Mn$_2$P$_2$S$_3$Se$_3$, with a Janus structure, exhibits the symmetries in this group, a topic that will be further discussed in the following section. In summary, we have derived spin layer groups above by combining spin group and layer groups to describe 2D  altermagnetism, and have provided their corresponding layer groups. In the subsequent section, we will give some examples of 2D  altermagnet and discuss their property of spin-momentum locking properties.

\begin{figure*}
	
	\centering
	\includegraphics[width=\linewidth]{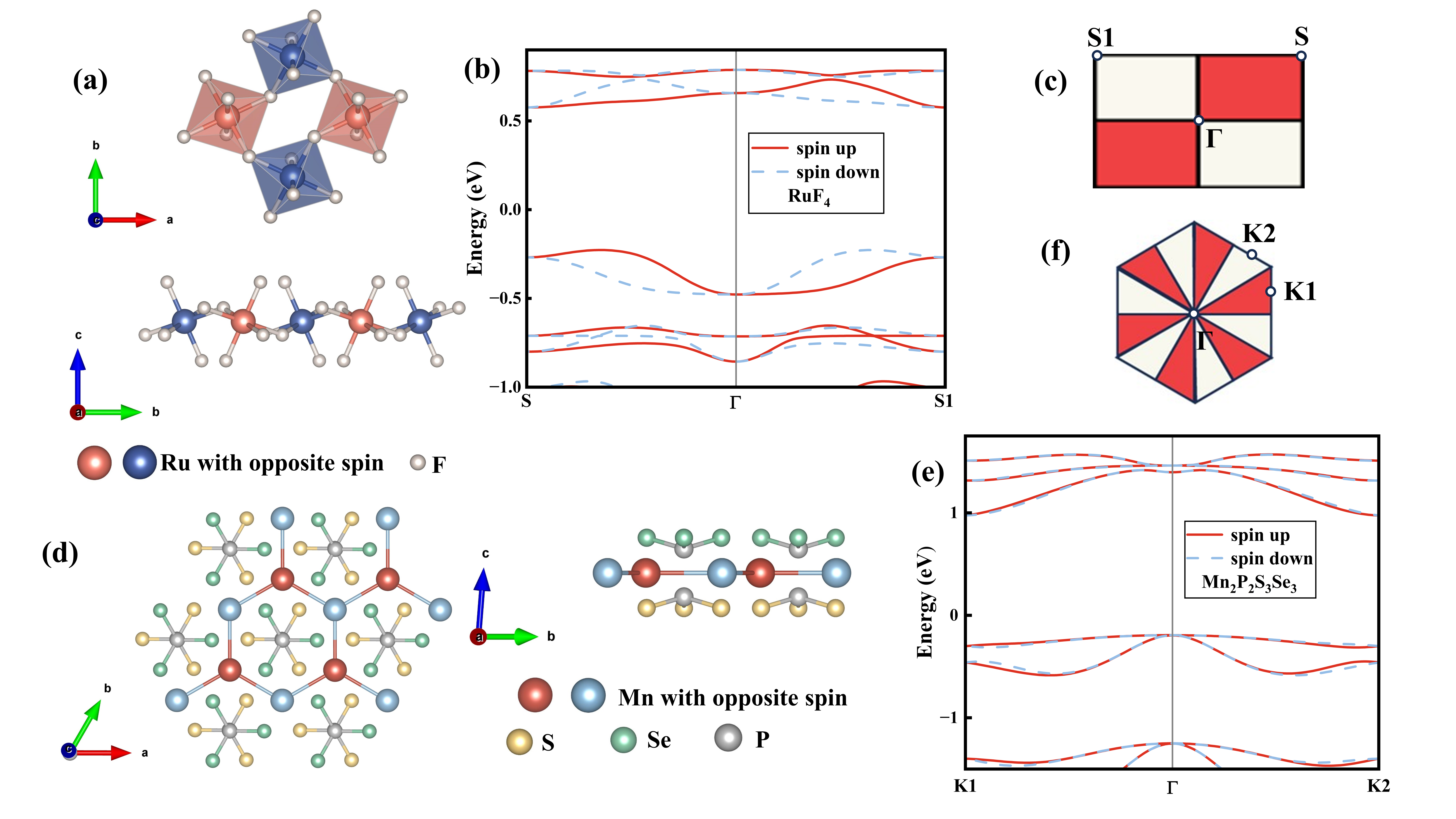}
	\caption{\label{fig:figure1}The crystal structure and nonrelativistic band structure of RuF$_4$ and Mn$_2$P$_2$S$_3$Se$_3$. (a), (d) The crystal structure of RuF$_4$ and Mn$_2$P$_2$S$_3$Se$_3$, where different colors of magnetic atom represent the opposite spin sublattices. (b), (e) The band structure of RuF$_4$ and Mn$_2$P$_2$S$_3$Se$_3$ without SOC, where red solid line and blue dotted line represent the opposite spin channels. (c), (f) The \textbf{k}-path we use to calculate band structure. The different colors represent opposite spins.}
	
\end{figure*}

\section{\label{sec: level 3}SPIN-MOMENTUM LOCKING IN MATERIAL CANDIDATE}

\begin{figure*}
	
	\centering
	\includegraphics[width=\linewidth]{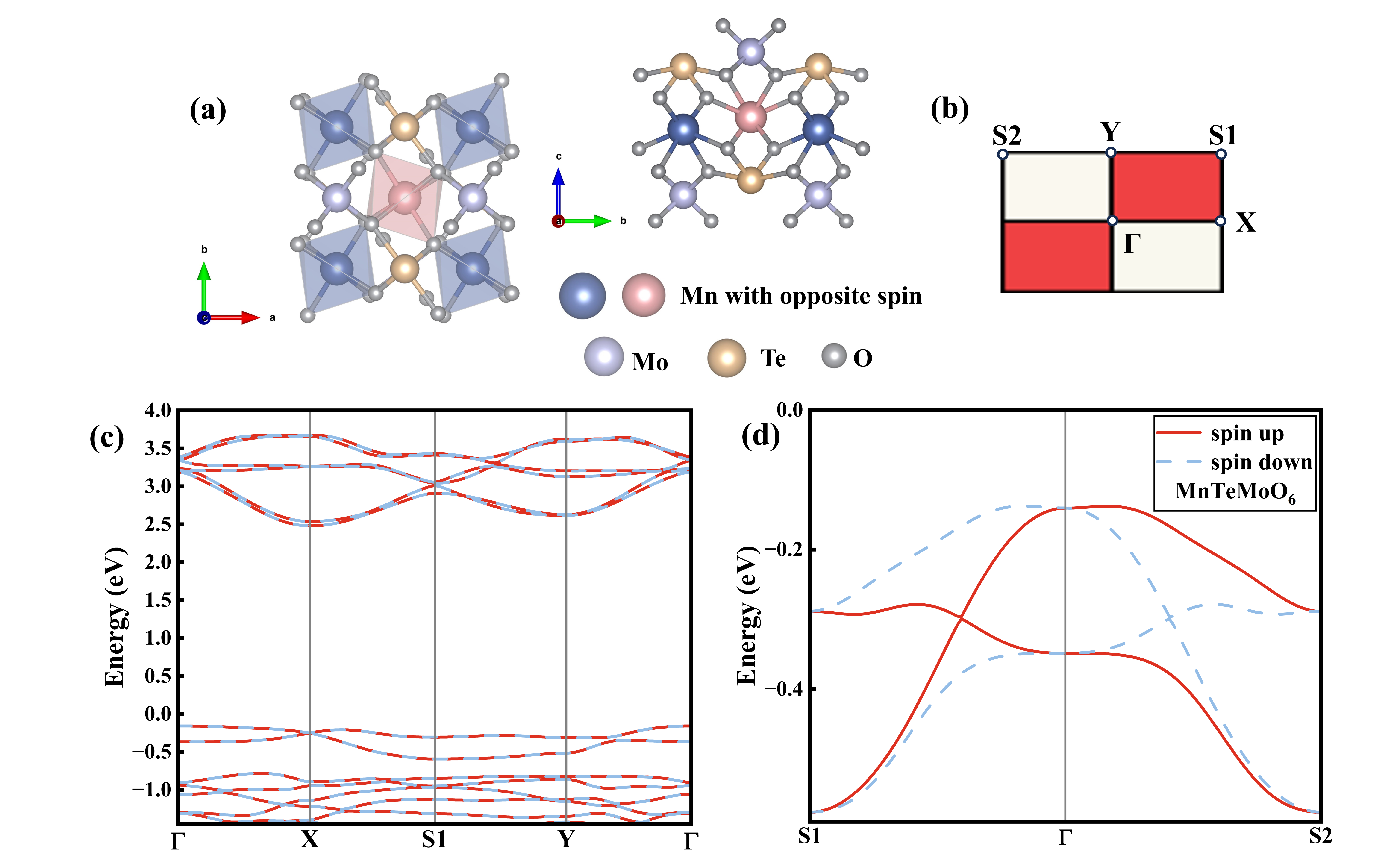}
	\caption{\label{fig:figure2}The crystal structure and nonrelativistic band structure of MnTeMoO$_6$. (a) The crystal structure of MnTeMoO$_6$, where pick and blue colors represent the opposite spin sublattices. (b) The Brillouin zone of MnTeMoO$_6$, in which the different colors represent opposite spins, and high-symmetry points used to calculate band structure.(c) The band structure of MnTeMoO$_6$ without SOC along \textbf{k}-path $\Gamma-X-S1-Y-\Gamma$ is spin-degeneracy, where red solid lines and blue dotted lines represent the opposite spin channels. (d) The band structure of MnTeMoO$_6$ without SOC along \textbf{k}-path $S1-\Gamma-S2$ is spin-splitting, where red solid lines and blue dotted lines represent the opposite spin channels.(c) and (d) clearly exhibit the basic characteristic of spin-momentum locking of spin layer group $^{2}m^{2}m^{1}m$. }
	
\end{figure*}

We now focus on the main characteristic of spin-momentum locking in 2D altermagnets, and subsequently, we will provide examples of the above derived spin layer groups of altermagnetism and discuss their characteristic of spin-momentum locking. Symmetry enables us to obtain information about band structure, such as spin-degeneracy \textbf{k}-vector and nodal line, prior to calculating the band structure. First of all, the spin-degeneracy nodal line is protected by the symmetry $[C_2||G-H]$ in $R_3$, particularly when this symmetry transforms the wave vector on this line to itself, i.e., $[C_2||G-H]\varepsilon(s,\textbf{k})=\varepsilon(-s,\textbf{k})=\varepsilon(s,\textbf{k})$. For instance, the symmetry $[C_2 ||C_{2x}]$, where $C_{2x}$ is a 2-fold rotation around the \emph{x}-axis, determines a spin-degeneracy nodal line along \emph{x}-axis. Moreover, if a symmetry $[C_2 ||G-H]$ in $R_3$ transforms a wave vector to itself or another wave vector separated by a reciprocal lattice vector, the band structure at this \textbf{k}-vector will exhibit spin-degeneracy. A typical example is the band structure at $\Gamma$-point, which is consistently spin-degenerate because any symmetry of $[C_2 ||G-H]$ transforms $\Gamma$-point to itself. Meanwhile, opposite spins will appear at two distinct \textbf{k}-vector which can be connected by the symmetry of $G-H$. This occurs due to  $[C_2||G-H]\varepsilon(s,\textbf{k})=\varepsilon(-s,\textbf{k}')$, where $\textbf{k}$ can be altered to $\textbf{k}'$ by the symmetry in $G-H$. This outcome is instrumental for determining the opposite spin-splitting $\textbf{k}$-vectors during the calculation of the band structure. The fundamental characteristic of spin-momentum locking for each spin layer group is listed  in the second column of the Table \ref{tab:table2}.

\begin{figure*}
	
	\centering
	\includegraphics[width=\linewidth]{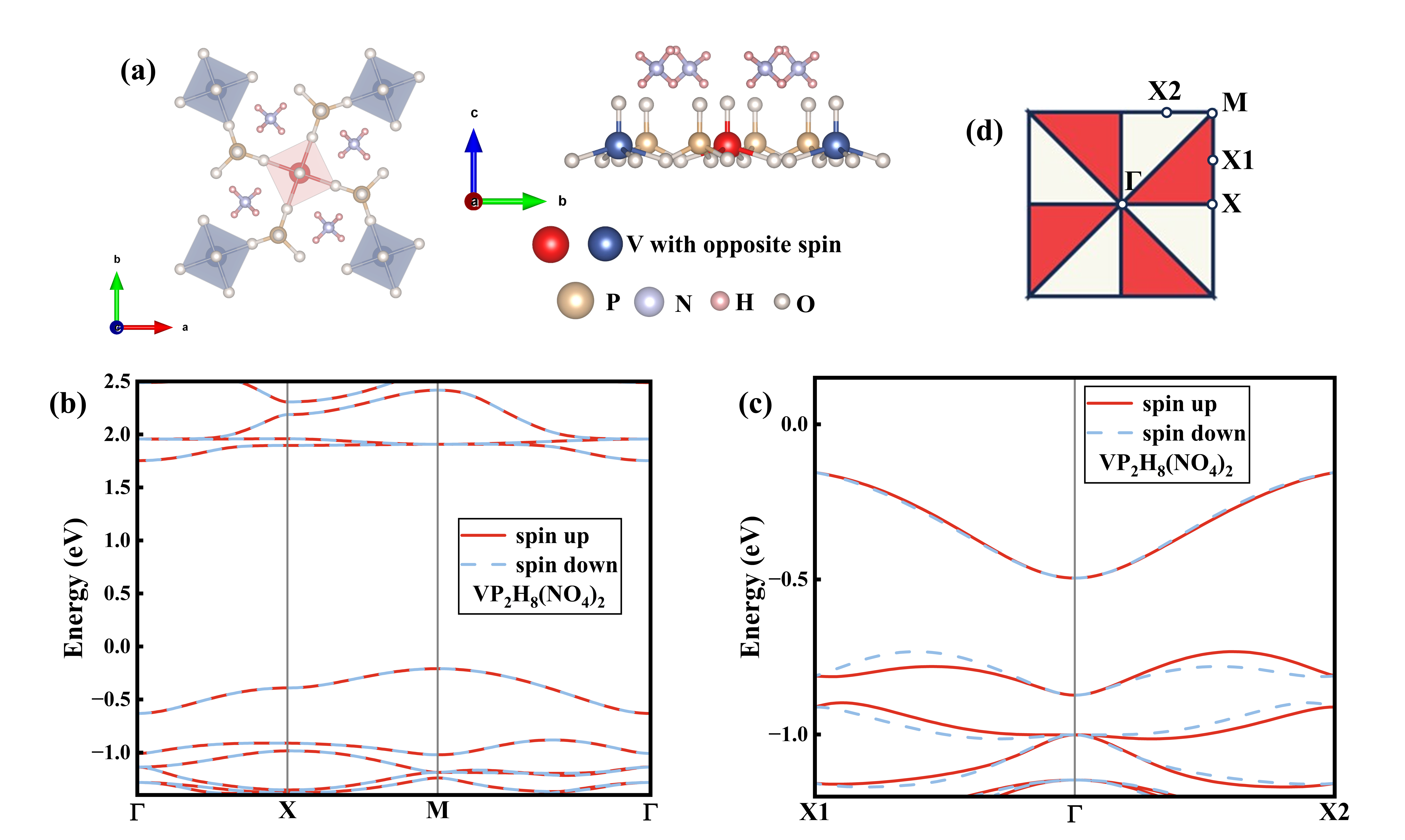}
	\caption{\label{fig:figure3}The crystal structure and nonrelativistic band structure of VP$_2$H$_8$(NO$_4$)$_2$. (a) The crystal structure of VP$_2$H$_8$(NO$_4$)$_2$, where red and blue colors represent the opposite spin sublattices. (b) The band structure of VP$_2$H$_8$(NO$_4$)$_2$ without SOC along \textbf{k}-path $\Gamma-X-M-\Gamma$ is spin-degeneracy, where red solid lines and blue dotted lines represent the opposite spin channels. (c) The band structure of VP$_2$H$_8$(NO$_4$)$_2$ without SOC along \textbf{k}-path $X1-\Gamma-X2$ is spin-splitting, where red solid lines and blue dotted lines represent the opposite spin channels.(b) and (c) clearly exhibit the basic characteristic of spin-momentum locking of spin layer group $^{1}4/^{1}m^{2}m^{2}m$.  (d) The Brillouin zone of VP$_2$H$_8$(NO$_4$)$_2$, in which the different colors represent opposite spins, and high-symmetry points used to calculate band structure. }
	
\end{figure*}
On the basis of spin layer group, we classify the reported 2D altermagnets, as listed in Table \ref{tab:table2}. It is observed that these materials mostly correspond to the spin layer group $^24 /^1m^2m^1m$. Furthermore, there are materials that exemplify two unexpected spin layer groups, which serve to validate our aforementioned results. RuF$_4$ is composed of Ru atoms located at the center of an octahedron formed by F atoms, as shown in FIG.2(a). Its layer group is NO.18, which contains the identity operation $E$, a 2-fold rotation symmetry around \emph{x}-axis $C_{2x}$, mirror symmetry through \emph{yz}-plane $m_x$ and real-space inversion symmetry $\overline{E}$. We also provide the magnetic configuration in altermagnetic state in FIG.\ref{fig:figure1}(a). Through symmetry analysis, it is demonstrated that atoms with opposite magnetic moment can be connected by $C_{2x}\tau$ and $m_x\tau$, where $\tau$ is a translation by half the lattice vector. Additionally,  atoms with same magnetic moment are connected by space-inversion symmetry. Consequently, its spin layer group is identified as  $^22/^2m_x$ and the \textbf{k}-path $\Gamma$-S exhibits an opposite spin sign to $\Gamma$-S1. The employed $\textbf{k}$-path is displayed in FIG.\ref{fig:figure1}(c). FIG.\ref{fig:figure1}(b) provides the nonrelativistic band structure of RuF$_4$, which corroborates the analysis results mentioned above.

The monolayer Mn$_2$P$_2$S$_3$Se$_3$ with a Janus structure has been identified as 2D altermagnet in previous report\cite{mazin2023induced}. Within the framework of spin layer group, it can be classified as $^{1}\overline{3}$$^2m$. In FIG.\ref{fig:figure1}(d), we illustrate the crystal structure of Mn$_2$P$_2$S$_3$Se$_3$, where all S atoms are below Mn plane and all Se atoms above, along with the magnetic configuration in altermagnetic state. Through symmetry analysis, it is revealed that Mn$_2$P$_2$S$_3$Se$_3$ belongs to layer group NO.70, which contains 3-fold rotation symmetry around \emph{z}-axis and three mirror symmetry through the planes perpendicular to \emph{xy}-plane, and atoms with opposite magnetic moments can be connected by mirror symmetry through a plane perpendicular to the \emph{xy}-plane. Therefore, its spin-layer-group is $^{1}\overline{3}$$^2m$ and the \textbf{k}-path $\Gamma$-K1 exhibits an opposite spin sign to $\Gamma$-K2. The $\textbf{k}$-path utilized in our analysis is shown in FIG.\ref{fig:figure1}(f). FIG.\ref{fig:figure1}(e) provides the band structure of Mn$_2$P$_2$S$_3$Se$_3$ which confirms the basic characteristic of spin-momentum locking with the spin layer group $^{1}\overline{3}$$^2m$.

However, four spin layer groups lack corresponding materials. We employ the symmetry operation in spin layer group to search for material candidates of these four groups in \emph{Material Project} database. For a specific spin layer group, we first selected materials containing transition metal elements and belonging to space groups that correspond to this spin layer group\cite{nespolo2017international,litvin2012character} within the \emph{Material Project} database. Then, we screened these materials individually. If a material with a vdW structure have an AFM magnetic configuration, we performed a symmetry analysis to determine whether it is altermagnet or not. Once we identified a material candidate for a spin layer group, we ceased our search for that group and initiated the search for the next spin layer group using the aforementioned process. This is because our aim was to identify material candidate for each spin layer group, rather than searching for all 2D altermagnets in \emph{Material Project} database. All of these operations were performed manually, without using any algorithms. It is suggested that monolayer MnTeMoO$_6$ has spin layer group $^2m^2m^1m$ and VP$_2$H$_8$(NO$_4$)$_2$ has $^14/^1m^2m^2m$; these are not reported in previous researches. Nevertheless, we have not identified the potential materials for the spin layer group $^24/^1m$ and $^16/^1m^2m^2m$, likely due to their high symmetry requirements. Of course, we believe that these two spin layer groups exist, as they are also represented within spin Laue group.

The crystal structure of MnTeMoO$_6$ is schematically illustrated in FIG.\ref{fig:figure2}(a), where pink and blue colors represent the real-space sublattices with opposite spins in altermagnetism. As a van der Waals material, it can be exfoliated from the bulk compound. Our DFT calculations reveal that the energy of the altermagnetic state is 8.5 meV per magnetic atom lower than that of the ferromagnetic state. Therefore, its magnetic ground state is the altermagnetic state. It has layer group NO.21, which contains 2-fold rotation symmetry around \emph{z}-axis and  two screw rotations around \emph{x}-axis and \emph{y}-axis respectively, and a spin layer group of $^2m^2m^1m$. The sublattices can be related by a 2-fold rotation around \emph{x}-axis (or \emph{y}-axis) with translation, and the 2-fold rotation around \emph{z}-axis maps the sublattice to itself. Therefore, the spin-momentum locking of MnTeMoO$_6$ is determined by the symmetry of $[C_2||C_{2x}]$ ($[C_2||C_{2y}]$), i.e., the spin-degeneracy nodal line at $k_x=0$ ($k_y=0$) is protected by $[C_2||C_{2x}]$ ($[C_2||C_{2y}]$). The \textbf{k}-paths, related by $C_{2x}$ ($C_{2y}$) but not separated by a reciprocal lattice vector, have opposite spin signs. FIG.\ref{fig:figure2}(c) and \ref{fig:figure2}(d) show the DFT calculated nonrelativistic band structure and the employed \textbf{k}-paths is shown in FIG.\ref{fig:figure2}(b). When we select a high-symmetry path along $\Gamma-X-S1-Y-\Gamma$ to calculate the band structure, we obtain only spin-degeneracy bands, as illustrated in FIG.\ref{fig:figure2}(c), which are protected by symmetry $[C_2||C_{2x}]$ ($[C_2||C_{2y}]$). However, it is observed that the band structure of \textbf{k}-path S1-$\Gamma$-S2 is spin splitting with the $\Gamma$-S1 path having an opposite spin to $\Gamma$-S2. This is a typical altermagnetic characteristic. These results are consistent with the conclusions we have reached earlier in our discussion.

For the spin layer group $^14/^1m^2m^2m$, we have identified VP$_2$H$_8$(NO$_4$)$_2$ as exhibiting this symmetry. It has a layer group NO.56, which contains a 4-fold rotation symmetry around \emph{z}-axis, a mirror symmetry through (110)plane and three glide reflections through \emph{xz}-plane, \emph{yz}-plane and ($1\overline{1}0$) plane. The crystal structure of VP$_2$H$_8$(NO$_4$)$_2$ is schematically illustrated in FIG.\ref{fig:figure3}(a), with red and blue colors representing the real-space sublattices with opposite spins in altermagnetism. Being a van der Waals material, it can be exfoliated from the bulk compound. We obtain that the energy of the altermagnetic state is 0.47 meV per magnetic atom lower than that of the ferromagnetic state by DFT calculations indicating that its magnetic ground state is an altermagnetic state. The sublattices can be related by reflection through the \emph{xz}-plane (\emph{yz}-plane, (110) or ($1\overline{1}0$) plane), and a 4-fold rotation around \emph{z}-axis maps sublattice to itself. Consequently, the spin-momentum locking of VP$_2$H$_8$(NO$_4$)$_2$ is determined by the symmetries $[C_2||\{m_x,m_y,m_{110},m_{1\overline{1}0}\}]$. That is the spin-degeneracy nodal line at $k_x=0$ and $k_y=0$ are protected by $[C_2||m_x]$ and $[C_2||m_y]$. The \textbf{k}-paths, related by $m_x$ ($m_y,m_{110},m_{1\overline{1}0}$) but not separated by a reciprocal lattice vector, have opposite spin signs. FIG.\ref{fig:figure3}(b) and (c) show the DFT calculated band structure without SOC and the employed \textbf{k}-paths is shown in FIG.\ref{fig:figure3}(d). If we choose high-symmetry path along $\Gamma-X-M-\Gamma$ to calculate the band structure, only the spin-degeneracy bands are obtained, as illustrated in FIG.\ref{fig:figure3}(b), which is protected by symmetry $m_x$ and $m_{1\overline{1}0}$. However, the band structure of \textbf{k}-path along X1-$\Gamma$-X2 is spin-spliting and $\Gamma$-X1 has opposite spin to $\Gamma$-X2, which is a typical altermagnetic characteristic. Again, these results are consistent with the conclusions we have reached above.

\section{CONCLUSION}
In summary, we have constructed seven spin layer groups to describe and classify 2D altermagnetism, extending beyond the previously reported plane-type spin Laue group. We believe that spin layer groups will be helpful in future search for 2D altermagnets. Meanwhile, we have utilized these spin layer groups to identify material candidates, validating the use of symmetry in the search for 2D altermagnets. We have also determined their magnetic ground state is altermagtic state and calculated the band structure by DFT calculations without SOC. Furthermore, we have conducted a symmetry-based analysis to evaluate the spin-momentum locking characteristics of the materials.
 
Nevertheless, several key issues require further investigation. Firstly, the significance of altermagnetism in 2D materials remains an open question, necessitating further discussion and DFT calculations with SOC. Secondly, there has been a lack of the experimental exploration into 2D altermagnets so far. Finally, determining whether the noncollinear magnetic state in altermagnetism differs from that in antiferromagnetism is a question that merits further exploration. 

\begin{acknowledgments}
	This work is financially supported by National Natural Science Foundation of China (Grant No. 12074126). The computer times at the High Performance Computational center at South China University of Technology are gratefully acknowledged. YJZ is grateful to the valuable suggestions provided by Prof. Junwei Liu.
\end{acknowledgments}

\appendix
\section{COMPUTATIONAL DETAILS}
All calculations were performed using the Vienna ab initio Simulation Package (VASP)\cite{kresse1996efficiency,kresse1996efficient},  employing the projector-augmented wave (PAW) method\cite{kresse1999ultrasoft} based on density functional theory. For the exchange-correlation functional, we use the generalized gradient approximation (GGA) with Perdew-Burke-Ernzerhof (PBE) functional\cite{perdew1996generalized}, along with Hubbard U correction\cite{PhysRevB.57.1505}. The employed U value and lattice constant are listed in Table \ref{tab:table3}. A cut-off energy of 500 eV was set for the plane wave basis. The structure was relaxed until the forces on atoms were below 0.01 eV/\AA\ and the convergence criteria was $1\times10^{-7}$ eV for the energy difference in electronic self-consistent calculation. A vacuum of 15\AA\ was constructed perpendicular to the material plane. The SOC effect was not considered  in calculations.
\begin{table}[h]
	\caption{\label{tab:table3}The employed U value and lattice constant. '--' denote that we do not apply Hubbard U correction on this material.}
	\begin{ruledtabular}
		\begin{tabular}{cccc}
			\textrm{material}&
			\textrm{U (eV)}&
			\textrm{\emph{a} (\AA)}&
			\textrm{\emph{b} (\AA)}\\
			\hline
			RuF$_4$ & --& 5.42 & 5.09 \\
			Mn$_2$P$_2$S$_3$Se$_3$ & -- & 6.22 & 6.22\\
			MnTeMoO$_6$ & 3.9(Mn), 4.38(Mo) & 5.14 & 5.41\\
			VP$_2$H$_8$(NO$_4$)$_2$ & 3.25(V) & 8.46 &8.46\\
		\end{tabular}
	\end{ruledtabular}
\end{table}

\begin{figure*}
	
	\centering
	\includegraphics[width=\linewidth]{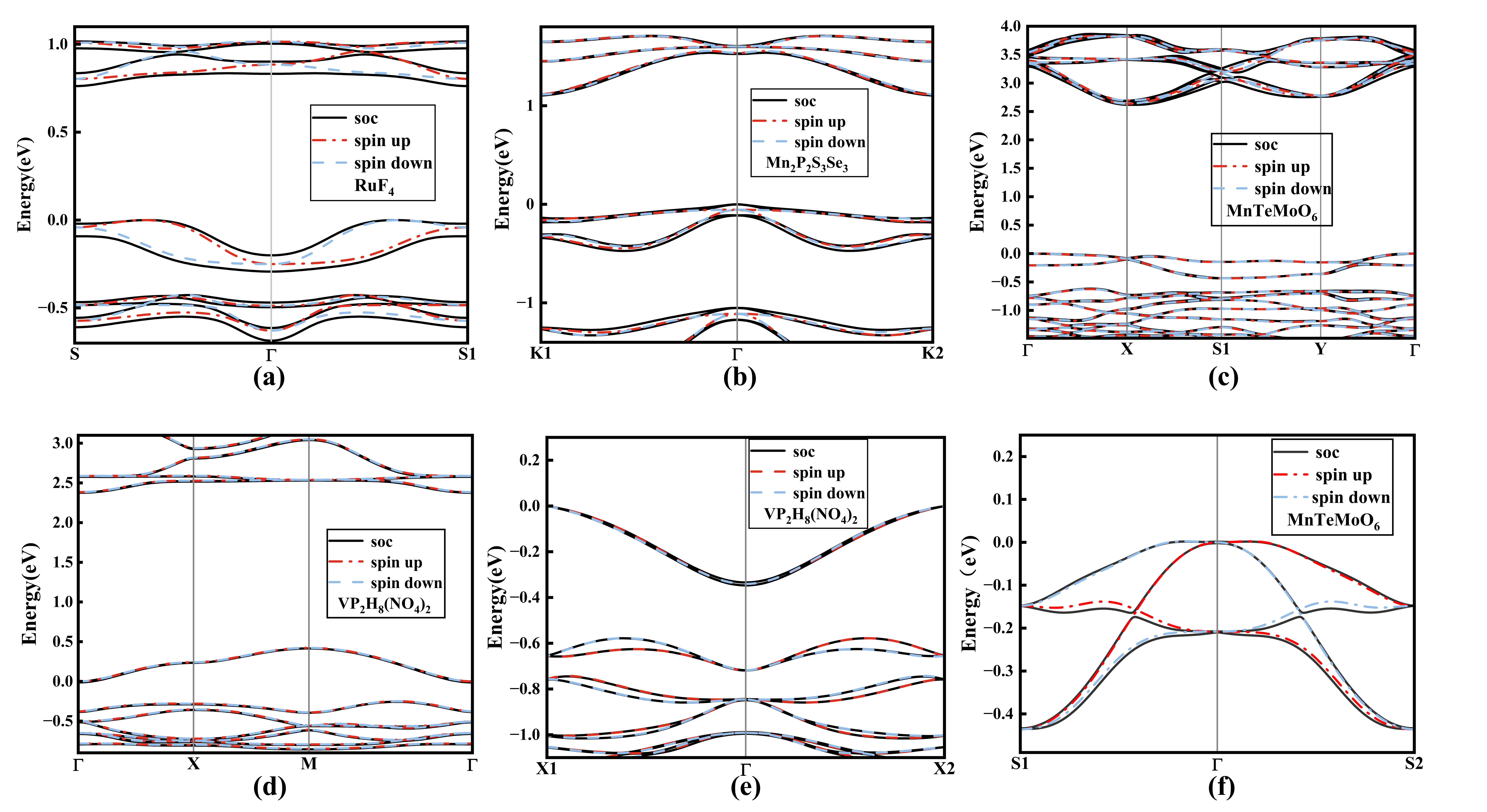}
	\caption{\label{fig:figure5}The band structure with and without SOC of (a)RuF$_4$, (b)Mn$_2$P$_2$S$_3$Se$_3$, (c)(f) MnTeMoO$_6$, and (d)(e)VP$_2$H$_8$(NO$_4$)$_2$. The red and blue dotted lines represent the opposite channels, and the black solid lines represent band structure with SOC. (a) corresponds to FIG.\ref{fig:figure1}(b). (b) corresponds to FIG.\ref{fig:figure1}(e). (c) and (f) corresponds to FIG.\ref{fig:figure2}(c) and (d). (d) and (e) corresponds to FIG.\ref{fig:figure3}(b) and (c).}
	
\end{figure*}

\section{COMPARISON OF BAND STRUCTURES WITH AND WITHOUT SOC}
As is known to all, the absence of SOC effect is an approximation. In this appendix, we provide comparison of band structures with and without SOC for all materials, as illustrated in FIG.\ref{fig:figure5}. It is implied that the magnitude of spin-splitting originating from SOC effect is small than that arises from altermagnetism. Therefore, altermagnetism plays a key role in spin-splitting in these materials and the discussion above without SOC is appropriate. It is worth noting that the intersections of two bands in FIG.\ref{fig:figure2}(d) vanish when we consider SOC effect, as illustrated in FIG.\ref{fig:figure5}(f). Consequently, we conclude that these points are not Weyl points.

\bibliography{ref}

\begin{thebibliography}{38}%
\makeatletter
\providecommand \@ifxundefined [1]{%
 \@ifx{#1\undefined}
}%
\providecommand \@ifnum [1]{%
 \ifnum #1\expandafter \@firstoftwo
 \else \expandafter \@secondoftwo
 \fi
}%
\providecommand \@ifx [1]{%
 \ifx #1\expandafter \@firstoftwo
 \else \expandafter \@secondoftwo
 \fi
}%
\providecommand \natexlab [1]{#1}%
\providecommand \enquote  [1]{``#1''}%
\providecommand \bibnamefont  [1]{#1}%
\providecommand \bibfnamefont [1]{#1}%
\providecommand \citenamefont [1]{#1}%
\providecommand \href@noop [0]{\@secondoftwo}%
\providecommand \href [0]{\begingroup \@sanitize@url \@href}%
\providecommand \@href[1]{\@@startlink{#1}\@@href}%
\providecommand \@@href[1]{\endgroup#1\@@endlink}%
\providecommand \@sanitize@url [0]{\catcode `\\12\catcode `\$12\catcode
  `\&12\catcode `\#12\catcode `\^12\catcode `\_12\catcode `\%12\relax}%
\providecommand \@@startlink[1]{}%
\providecommand \@@endlink[0]{}%
\providecommand \url  [0]{\begingroup\@sanitize@url \@url }%
\providecommand \@url [1]{\endgroup\@href {#1}{\urlprefix }}%
\providecommand \urlprefix  [0]{URL }%
\providecommand \Eprint [0]{\href }%
\providecommand \doibase [0]{https://doi.org/}%
\providecommand \selectlanguage [0]{\@gobble}%
\providecommand \bibinfo  [0]{\@secondoftwo}%
\providecommand \bibfield  [0]{\@secondoftwo}%
\providecommand \translation [1]{[#1]}%
\providecommand \BibitemOpen [0]{}%
\providecommand \bibitemStop [0]{}%
\providecommand \bibitemNoStop [0]{.\EOS\space}%
\providecommand \EOS [0]{\spacefactor3000\relax}%
\providecommand \BibitemShut  [1]{\csname bibitem#1\endcsname}%
\let\auto@bib@innerbib\@empty
\bibitem [{\citenamefont {\ifmmode~\check{S}\else \v{S}\fi{}mejkal}\ \emph
  {et~al.}(2022{\natexlab{a}})\citenamefont {\ifmmode~\check{S}\else
  \v{S}\fi{}mejkal}, \citenamefont {Sinova},\ and\ \citenamefont
  {Jungwirth}}]{vsmejkal2022emerging}%
  \BibitemOpen
  \bibfield  {author} {\bibinfo {author} {\bibfnamefont {L.}~\bibnamefont
  {\ifmmode~\check{S}\else \v{S}\fi{}mejkal}}, \bibinfo {author} {\bibfnamefont
  {J.}~\bibnamefont {Sinova}},\ and\ \bibinfo {author} {\bibfnamefont
  {T.}~\bibnamefont {Jungwirth}},\ }\bibfield  {title} {\bibinfo {title}
  {Emerging research landscape of altermagnetism},\ }\href
  {https://doi.org/10.1103/PhysRevX.12.040501} {\bibfield  {journal} {\bibinfo
  {journal} {Phys. Rev. X}\ }\textbf {\bibinfo {volume} {12}},\ \bibinfo
  {pages} {040501} (\bibinfo {year} {2022}{\natexlab{a}})}\BibitemShut
  {NoStop}%
\bibitem [{\citenamefont {Zhou}\ \emph {et~al.}(2024)\citenamefont {Zhou},
  \citenamefont {Feng}, \citenamefont {Zhang}, \citenamefont
  {\ifmmode~\check{S}\else \v{S}\fi{}mejkal}, \citenamefont {Sinova},
  \citenamefont {Mokrousov},\ and\ \citenamefont {Yao}}]{zhou2024crystal}%
  \BibitemOpen
  \bibfield  {author} {\bibinfo {author} {\bibfnamefont {X.}~\bibnamefont
  {Zhou}}, \bibinfo {author} {\bibfnamefont {W.}~\bibnamefont {Feng}}, \bibinfo
  {author} {\bibfnamefont {R.-W.}\ \bibnamefont {Zhang}}, \bibinfo {author}
  {\bibfnamefont {L.}~\bibnamefont {\ifmmode~\check{S}\else \v{S}\fi{}mejkal}},
  \bibinfo {author} {\bibfnamefont {J.}~\bibnamefont {Sinova}}, \bibinfo
  {author} {\bibfnamefont {Y.}~\bibnamefont {Mokrousov}},\ and\ \bibinfo
  {author} {\bibfnamefont {Y.}~\bibnamefont {Yao}},\ }\bibfield  {title}
  {\bibinfo {title} {Crystal thermal transport in altermagnetic
  {${\mathrm{RuO}}_{2}$}},\ }\href
  {https://doi.org/10.1103/PhysRevLett.132.056701} {\bibfield  {journal}
  {\bibinfo  {journal} {Phys. Rev. Lett.}\ }\textbf {\bibinfo {volume} {132}},\
  \bibinfo {pages} {056701} (\bibinfo {year} {2024})}\BibitemShut {NoStop}%
\bibitem [{\citenamefont {Mazin}(2024)}]{mazin2024altermagnetism}%
  \BibitemOpen
  \bibfield  {author} {\bibinfo {author} {\bibfnamefont {I.}~\bibnamefont
  {Mazin}},\ }\bibfield  {title} {\bibinfo {title} {Altermagnetism then and
  now},\ }\href@noop {} {\bibfield  {journal} {\bibinfo  {journal} {Physics}\
  }\textbf {\bibinfo {volume} {17}},\ \bibinfo {pages} {4} (\bibinfo {year}
  {2024})}\BibitemShut {NoStop}%
\bibitem [{\citenamefont {Chen}\ \emph
  {et~al.}(2023{\natexlab{a}})\citenamefont {Chen}, \citenamefont {Ren},
  \citenamefont {Li}, \citenamefont {Liu},\ and\ \citenamefont
  {Liu}}]{chen2023spin}%
  \BibitemOpen
  \bibfield  {author} {\bibinfo {author} {\bibfnamefont {X.}~\bibnamefont
  {Chen}}, \bibinfo {author} {\bibfnamefont {J.}~\bibnamefont {Ren}}, \bibinfo
  {author} {\bibfnamefont {J.}~\bibnamefont {Li}}, \bibinfo {author}
  {\bibfnamefont {Y.}~\bibnamefont {Liu}},\ and\ \bibinfo {author}
  {\bibfnamefont {Q.}~\bibnamefont {Liu}},\ }\bibfield  {title} {\bibinfo
  {title} {Spin space group theory and unconventional magnons in collinear
  magnets},\ }\href@noop {} {\bibfield  {journal} {\bibinfo  {journal} {arXiv
  preprint arXiv:2307.12366}\ } (\bibinfo {year}
  {2023}{\natexlab{a}})}\BibitemShut {NoStop}%
\bibitem [{\citenamefont {Ma}\ \emph {et~al.}(2021)\citenamefont {Ma},
  \citenamefont {Hu}, \citenamefont {Li}, \citenamefont {Liu}, \citenamefont
  {Yao}, \citenamefont {Jia},\ and\ \citenamefont
  {Liu}}]{ma2021multifunctional}%
  \BibitemOpen
  \bibfield  {author} {\bibinfo {author} {\bibfnamefont {H.-Y.}\ \bibnamefont
  {Ma}}, \bibinfo {author} {\bibfnamefont {M.}~\bibnamefont {Hu}}, \bibinfo
  {author} {\bibfnamefont {N.}~\bibnamefont {Li}}, \bibinfo {author}
  {\bibfnamefont {J.}~\bibnamefont {Liu}}, \bibinfo {author} {\bibfnamefont
  {W.}~\bibnamefont {Yao}}, \bibinfo {author} {\bibfnamefont {J.-F.}\
  \bibnamefont {Jia}},\ and\ \bibinfo {author} {\bibfnamefont {J.}~\bibnamefont
  {Liu}},\ }\bibfield  {title} {\bibinfo {title} {Multifunctional
  antiferromagnetic materials with giant piezomagnetism and noncollinear spin
  current},\ }\href@noop {} {\bibfield  {journal} {\bibinfo  {journal} {Nature
  communications}\ }\textbf {\bibinfo {volume} {12}},\ \bibinfo {pages} {2846}
  (\bibinfo {year} {2021})}\BibitemShut {NoStop}%
\bibitem [{\citenamefont {Mazin}\ \emph {et~al.}(2021)\citenamefont {Mazin},
  \citenamefont {Koepernik}, \citenamefont {Johannes}, \citenamefont
  {Gonz{\'a}lez-Hern{\'a}ndez},\ and\ \citenamefont
  {{\v{S}}mejkal}}]{mazin2021prediction}%
  \BibitemOpen
  \bibfield  {author} {\bibinfo {author} {\bibfnamefont {I.~I.}\ \bibnamefont
  {Mazin}}, \bibinfo {author} {\bibfnamefont {K.}~\bibnamefont {Koepernik}},
  \bibinfo {author} {\bibfnamefont {M.~D.}\ \bibnamefont {Johannes}}, \bibinfo
  {author} {\bibfnamefont {R.}~\bibnamefont {Gonz{\'a}lez-Hern{\'a}ndez}},\
  and\ \bibinfo {author} {\bibfnamefont {L.}~\bibnamefont {{\v{S}}mejkal}},\
  }\bibfield  {title} {\bibinfo {title} {Prediction of unconventional magnetism
  in doped {FeSb$_2$}},\ }\href@noop {} {\bibfield  {journal} {\bibinfo
  {journal} {Proceedings of the National Academy of Sciences}\ }\textbf
  {\bibinfo {volume} {118}},\ \bibinfo {pages} {e2108924118} (\bibinfo {year}
  {2021})}\BibitemShut {NoStop}%
\bibitem [{\citenamefont {Hayami}\ \emph {et~al.}(2019)\citenamefont {Hayami},
  \citenamefont {Yanagi},\ and\ \citenamefont {Kusunose}}]{hayami2019momentum}%
  \BibitemOpen
  \bibfield  {author} {\bibinfo {author} {\bibfnamefont {S.}~\bibnamefont
  {Hayami}}, \bibinfo {author} {\bibfnamefont {Y.}~\bibnamefont {Yanagi}},\
  and\ \bibinfo {author} {\bibfnamefont {H.}~\bibnamefont {Kusunose}},\
  }\bibfield  {title} {\bibinfo {title} {Momentum-dependent spin splitting by
  collinear antiferromagnetic ordering},\ }\href@noop {} {\bibfield  {journal}
  {\bibinfo  {journal} {journal of the physical society of japan}\ }\textbf
  {\bibinfo {volume} {88}},\ \bibinfo {pages} {123702} (\bibinfo {year}
  {2019})}\BibitemShut {NoStop}%
\bibitem [{\citenamefont {{\v{S}}mejkal}\ \emph {et~al.}(2020)\citenamefont
  {{\v{S}}mejkal}, \citenamefont {Gonz{\'a}lez-Hern{\'a}ndez}, \citenamefont
  {Jungwirth},\ and\ \citenamefont {Sinova}}]{vsmejkal2020crystal}%
  \BibitemOpen
  \bibfield  {author} {\bibinfo {author} {\bibfnamefont {L.}~\bibnamefont
  {{\v{S}}mejkal}}, \bibinfo {author} {\bibfnamefont {R.}~\bibnamefont
  {Gonz{\'a}lez-Hern{\'a}ndez}}, \bibinfo {author} {\bibfnamefont
  {T.}~\bibnamefont {Jungwirth}},\ and\ \bibinfo {author} {\bibfnamefont
  {J.}~\bibnamefont {Sinova}},\ }\bibfield  {title} {\bibinfo {title} {Crystal
  time-reversal symmetry breaking and spontaneous hall effect in collinear
  antiferromagnets},\ }\href@noop {} {\bibfield  {journal} {\bibinfo  {journal}
  {Science advances}\ }\textbf {\bibinfo {volume} {6}},\ \bibinfo {pages}
  {eaaz8809} (\bibinfo {year} {2020})}\BibitemShut {NoStop}%
\bibitem [{\citenamefont {Yuan}\ \emph {et~al.}(2020)\citenamefont {Yuan},
  \citenamefont {Wang}, \citenamefont {Luo}, \citenamefont {Rashba},\ and\
  \citenamefont {Zunger}}]{PhysRevB.102.014422}%
  \BibitemOpen
  \bibfield  {author} {\bibinfo {author} {\bibfnamefont {L.-D.}\ \bibnamefont
  {Yuan}}, \bibinfo {author} {\bibfnamefont {Z.}~\bibnamefont {Wang}}, \bibinfo
  {author} {\bibfnamefont {J.-W.}\ \bibnamefont {Luo}}, \bibinfo {author}
  {\bibfnamefont {E.~I.}\ \bibnamefont {Rashba}},\ and\ \bibinfo {author}
  {\bibfnamefont {A.}~\bibnamefont {Zunger}},\ }\bibfield  {title} {\bibinfo
  {title} {Giant momentum-dependent spin splitting in centrosymmetric low-{$Z$}
  antiferromagnets},\ }\href {https://doi.org/10.1103/PhysRevB.102.014422}
  {\bibfield  {journal} {\bibinfo  {journal} {Phys. Rev. B}\ }\textbf {\bibinfo
  {volume} {102}},\ \bibinfo {pages} {014422} (\bibinfo {year}
  {2020})}\BibitemShut {NoStop}%
\bibitem [{\citenamefont {\ifmmode~\check{S}\else \v{S}\fi{}mejkal}\ \emph
  {et~al.}(2022{\natexlab{b}})\citenamefont {\ifmmode~\check{S}\else
  \v{S}\fi{}mejkal}, \citenamefont {Sinova},\ and\ \citenamefont
  {Jungwirth}}]{vsmejkal2022beyond}%
  \BibitemOpen
  \bibfield  {author} {\bibinfo {author} {\bibfnamefont {L.}~\bibnamefont
  {\ifmmode~\check{S}\else \v{S}\fi{}mejkal}}, \bibinfo {author} {\bibfnamefont
  {J.}~\bibnamefont {Sinova}},\ and\ \bibinfo {author} {\bibfnamefont
  {T.}~\bibnamefont {Jungwirth}},\ }\bibfield  {title} {\bibinfo {title}
  {Beyond conventional ferromagnetism and antiferromagnetism: A phase with
  nonrelativistic spin and crystal rotation symmetry},\ }\href
  {https://doi.org/10.1103/PhysRevX.12.031042} {\bibfield  {journal} {\bibinfo
  {journal} {Phys. Rev. X}\ }\textbf {\bibinfo {volume} {12}},\ \bibinfo
  {pages} {031042} (\bibinfo {year} {2022}{\natexlab{b}})}\BibitemShut
  {NoStop}%
\bibitem [{\citenamefont {Brinkman}\ and\ \citenamefont
  {Elliott}(1966)}]{brinkman1966theory}%
  \BibitemOpen
  \bibfield  {author} {\bibinfo {author} {\bibfnamefont {W.}~\bibnamefont
  {Brinkman}}\ and\ \bibinfo {author} {\bibfnamefont {R.~J.}\ \bibnamefont
  {Elliott}},\ }\bibfield  {title} {\bibinfo {title} {Theory of spin-space
  groups},\ }\href@noop {} {\bibfield  {journal} {\bibinfo  {journal}
  {Proceedings of the Royal Society of London. Series A. Mathematical and
  Physical Sciences}\ }\textbf {\bibinfo {volume} {294}},\ \bibinfo {pages}
  {343} (\bibinfo {year} {1966})}\BibitemShut {NoStop}%
\bibitem [{\citenamefont {Litvin}\ and\ \citenamefont
  {Opechowski}(1974)}]{litvin1974spin}%
  \BibitemOpen
  \bibfield  {author} {\bibinfo {author} {\bibfnamefont {D.~B.}\ \bibnamefont
  {Litvin}}\ and\ \bibinfo {author} {\bibfnamefont {W.}~\bibnamefont
  {Opechowski}},\ }\bibfield  {title} {\bibinfo {title} {Spin groups},\
  }\href@noop {} {\bibfield  {journal} {\bibinfo  {journal} {Physica}\ }\textbf
  {\bibinfo {volume} {76}},\ \bibinfo {pages} {538} (\bibinfo {year}
  {1974})}\BibitemShut {NoStop}%
\bibitem [{\citenamefont {Litvin}(1977)}]{litvin1977spin}%
  \BibitemOpen
  \bibfield  {author} {\bibinfo {author} {\bibfnamefont {D.~B.}\ \bibnamefont
  {Litvin}},\ }\bibfield  {title} {\bibinfo {title} {Spin point groups},\
  }\href@noop {} {\bibfield  {journal} {\bibinfo  {journal} {Acta
  Crystallographica Section A: Crystal Physics, Diffraction, Theoretical and
  General Crystallography}\ }\textbf {\bibinfo {volume} {33}},\ \bibinfo
  {pages} {279} (\bibinfo {year} {1977})}\BibitemShut {NoStop}%
\bibitem [{\citenamefont {{\v{S}}mejkal}\ \emph {et~al.}(2022)\citenamefont
  {{\v{S}}mejkal}, \citenamefont {MacDonald}, \citenamefont {Sinova},
  \citenamefont {Nakatsuji},\ and\ \citenamefont
  {Jungwirth}}]{vsmejkal2022anomalous}%
  \BibitemOpen
  \bibfield  {author} {\bibinfo {author} {\bibfnamefont {L.}~\bibnamefont
  {{\v{S}}mejkal}}, \bibinfo {author} {\bibfnamefont {A.~H.}\ \bibnamefont
  {MacDonald}}, \bibinfo {author} {\bibfnamefont {J.}~\bibnamefont {Sinova}},
  \bibinfo {author} {\bibfnamefont {S.}~\bibnamefont {Nakatsuji}},\ and\
  \bibinfo {author} {\bibfnamefont {T.}~\bibnamefont {Jungwirth}},\ }\bibfield
  {title} {\bibinfo {title} {Anomalous hall antiferromagnets},\ }\href@noop {}
  {\bibfield  {journal} {\bibinfo  {journal} {Nature Reviews Materials}\
  }\textbf {\bibinfo {volume} {7}},\ \bibinfo {pages} {482} (\bibinfo {year}
  {2022})}\BibitemShut {NoStop}%
\bibitem [{\citenamefont {Samanta}\ \emph {et~al.}(2020)\citenamefont
  {Samanta}, \citenamefont {Le{\v{z}}ai{\'c}}, \citenamefont {Merte},
  \citenamefont {Freimuth}, \citenamefont {Bl{\"u}gel},\ and\ \citenamefont
  {Mokrousov}}]{samanta2020crystal}%
  \BibitemOpen
  \bibfield  {author} {\bibinfo {author} {\bibfnamefont {K.}~\bibnamefont
  {Samanta}}, \bibinfo {author} {\bibfnamefont {M.}~\bibnamefont
  {Le{\v{z}}ai{\'c}}}, \bibinfo {author} {\bibfnamefont {M.}~\bibnamefont
  {Merte}}, \bibinfo {author} {\bibfnamefont {F.}~\bibnamefont {Freimuth}},
  \bibinfo {author} {\bibfnamefont {S.}~\bibnamefont {Bl{\"u}gel}},\ and\
  \bibinfo {author} {\bibfnamefont {Y.}~\bibnamefont {Mokrousov}},\ }\bibfield
  {title} {\bibinfo {title} {Crystal hall and crystal magneto-optical effect in
  thin films of {${\mathrm{SrRuO}}_{3}$}},\ }\href@noop {} {\bibfield
  {journal} {\bibinfo  {journal} {Journal of applied physics}\ }\textbf
  {\bibinfo {volume} {127}} (\bibinfo {year} {2020})}\BibitemShut {NoStop}%
\bibitem [{\citenamefont {Zhou}\ \emph {et~al.}(2021)\citenamefont {Zhou},
  \citenamefont {Feng}, \citenamefont {Yang}, \citenamefont {Guo},\ and\
  \citenamefont {Yao}}]{zhou2021crystal}%
  \BibitemOpen
  \bibfield  {author} {\bibinfo {author} {\bibfnamefont {X.}~\bibnamefont
  {Zhou}}, \bibinfo {author} {\bibfnamefont {W.}~\bibnamefont {Feng}}, \bibinfo
  {author} {\bibfnamefont {X.}~\bibnamefont {Yang}}, \bibinfo {author}
  {\bibfnamefont {G.-Y.}\ \bibnamefont {Guo}},\ and\ \bibinfo {author}
  {\bibfnamefont {Y.}~\bibnamefont {Yao}},\ }\bibfield  {title} {\bibinfo
  {title} {Crystal chirality magneto-optical effects in collinear
  antiferromagnets},\ }\href {https://doi.org/10.1103/PhysRevB.104.024401}
  {\bibfield  {journal} {\bibinfo  {journal} {Phys. Rev. B}\ }\textbf {\bibinfo
  {volume} {104}},\ \bibinfo {pages} {024401} (\bibinfo {year}
  {2021})}\BibitemShut {NoStop}%
\bibitem [{\citenamefont {Bai}\ \emph {et~al.}(2022)\citenamefont {Bai},
  \citenamefont {Han}, \citenamefont {Feng}, \citenamefont {Zhou},
  \citenamefont {Su}, \citenamefont {Wang}, \citenamefont {Liao}, \citenamefont
  {Zhu}, \citenamefont {Chen}, \citenamefont {Pan}, \citenamefont {Fan},\ and\
  \citenamefont {Song}}]{bai2022observation}%
  \BibitemOpen
  \bibfield  {author} {\bibinfo {author} {\bibfnamefont {H.}~\bibnamefont
  {Bai}}, \bibinfo {author} {\bibfnamefont {L.}~\bibnamefont {Han}}, \bibinfo
  {author} {\bibfnamefont {X.~Y.}\ \bibnamefont {Feng}}, \bibinfo {author}
  {\bibfnamefont {Y.~J.}\ \bibnamefont {Zhou}}, \bibinfo {author}
  {\bibfnamefont {R.~X.}\ \bibnamefont {Su}}, \bibinfo {author} {\bibfnamefont
  {Q.}~\bibnamefont {Wang}}, \bibinfo {author} {\bibfnamefont {L.~Y.}\
  \bibnamefont {Liao}}, \bibinfo {author} {\bibfnamefont {W.~X.}\ \bibnamefont
  {Zhu}}, \bibinfo {author} {\bibfnamefont {X.~Z.}\ \bibnamefont {Chen}},
  \bibinfo {author} {\bibfnamefont {F.}~\bibnamefont {Pan}}, \bibinfo {author}
  {\bibfnamefont {X.~L.}\ \bibnamefont {Fan}},\ and\ \bibinfo {author}
  {\bibfnamefont {C.}~\bibnamefont {Song}},\ }\bibfield  {title} {\bibinfo
  {title} {Observation of spin splitting torque in a collinear antiferromagnet
  {${\mathrm{RuO}}_{2}$}},\ }\href
  {https://doi.org/10.1103/PhysRevLett.128.197202} {\bibfield  {journal}
  {\bibinfo  {journal} {Phys. Rev. Lett.}\ }\textbf {\bibinfo {volume} {128}},\
  \bibinfo {pages} {197202} (\bibinfo {year} {2022})}\BibitemShut {NoStop}%
\bibitem [{\citenamefont {Lee}\ \emph {et~al.}(2024)\citenamefont {Lee},
  \citenamefont {Lee}, \citenamefont {Jung}, \citenamefont {Jung},
  \citenamefont {Kim}, \citenamefont {Lee}, \citenamefont {Seok}, \citenamefont
  {Kim}, \citenamefont {Park}, \citenamefont {\ifmmode~\check{S}\else
  \v{S}\fi{}mejkal}, \citenamefont {Kang},\ and\ \citenamefont
  {Kim}}]{lee2024broken}%
  \BibitemOpen
  \bibfield  {author} {\bibinfo {author} {\bibfnamefont {S.}~\bibnamefont
  {Lee}}, \bibinfo {author} {\bibfnamefont {S.}~\bibnamefont {Lee}}, \bibinfo
  {author} {\bibfnamefont {S.}~\bibnamefont {Jung}}, \bibinfo {author}
  {\bibfnamefont {J.}~\bibnamefont {Jung}}, \bibinfo {author} {\bibfnamefont
  {D.}~\bibnamefont {Kim}}, \bibinfo {author} {\bibfnamefont {Y.}~\bibnamefont
  {Lee}}, \bibinfo {author} {\bibfnamefont {B.}~\bibnamefont {Seok}}, \bibinfo
  {author} {\bibfnamefont {J.}~\bibnamefont {Kim}}, \bibinfo {author}
  {\bibfnamefont {B.~G.}\ \bibnamefont {Park}}, \bibinfo {author}
  {\bibfnamefont {L.}~\bibnamefont {\ifmmode~\check{S}\else \v{S}\fi{}mejkal}},
  \bibinfo {author} {\bibfnamefont {C.-J.}\ \bibnamefont {Kang}},\ and\
  \bibinfo {author} {\bibfnamefont {C.}~\bibnamefont {Kim}},\ }\bibfield
  {title} {\bibinfo {title} {Broken kramers degeneracy in altermagnetic
  {MnTe}},\ }\href {https://doi.org/10.1103/PhysRevLett.132.036702} {\bibfield
  {journal} {\bibinfo  {journal} {Phys. Rev. Lett.}\ }\textbf {\bibinfo
  {volume} {132}},\ \bibinfo {pages} {036702} (\bibinfo {year}
  {2024})}\BibitemShut {NoStop}%
\bibitem [{\citenamefont {\ifmmode~\check{S}\else \v{S}\fi{}mejkal}\ \emph
  {et~al.}(2022{\natexlab{c}})\citenamefont {\ifmmode~\check{S}\else
  \v{S}\fi{}mejkal}, \citenamefont {Hellenes}, \citenamefont
  {Gonz\'alez-Hern\'andez}, \citenamefont {Sinova},\ and\ \citenamefont
  {Jungwirth}}]{PhysRevX.12.011028}%
  \BibitemOpen
  \bibfield  {author} {\bibinfo {author} {\bibfnamefont {L.}~\bibnamefont
  {\ifmmode~\check{S}\else \v{S}\fi{}mejkal}}, \bibinfo {author} {\bibfnamefont
  {A.~B.}\ \bibnamefont {Hellenes}}, \bibinfo {author} {\bibfnamefont
  {R.}~\bibnamefont {Gonz\'alez-Hern\'andez}}, \bibinfo {author} {\bibfnamefont
  {J.}~\bibnamefont {Sinova}},\ and\ \bibinfo {author} {\bibfnamefont
  {T.}~\bibnamefont {Jungwirth}},\ }\bibfield  {title} {\bibinfo {title} {Giant
  and tunneling magnetoresistance in unconventional collinear antiferromagnets
  with nonrelativistic spin-momentum coupling},\ }\href
  {https://doi.org/10.1103/PhysRevX.12.011028} {\bibfield  {journal} {\bibinfo
  {journal} {Phys. Rev. X}\ }\textbf {\bibinfo {volume} {12}},\ \bibinfo
  {pages} {011028} (\bibinfo {year} {2022}{\natexlab{c}})}\BibitemShut
  {NoStop}%
\bibitem [{\citenamefont {Gonz\'alez-Hern\'andez}\ \emph
  {et~al.}(2021)\citenamefont {Gonz\'alez-Hern\'andez}, \citenamefont
  {\ifmmode~\check{S}\else \v{S}\fi{}mejkal}, \citenamefont {V\'yborn\'y},
  \citenamefont {Yahagi}, \citenamefont {Sinova}, \citenamefont {Jungwirth},\
  and\ \citenamefont {\ifmmode~\check{Z}\else
  \v{Z}\fi{}elezn\'y}}]{PhysRevLett.126.127701}%
  \BibitemOpen
  \bibfield  {author} {\bibinfo {author} {\bibfnamefont {R.}~\bibnamefont
  {Gonz\'alez-Hern\'andez}}, \bibinfo {author} {\bibfnamefont {L.}~\bibnamefont
  {\ifmmode~\check{S}\else \v{S}\fi{}mejkal}}, \bibinfo {author} {\bibfnamefont
  {K.}~\bibnamefont {V\'yborn\'y}}, \bibinfo {author} {\bibfnamefont
  {Y.}~\bibnamefont {Yahagi}}, \bibinfo {author} {\bibfnamefont
  {J.}~\bibnamefont {Sinova}}, \bibinfo {author} {\bibfnamefont
  {T.}~\bibnamefont {Jungwirth}},\ and\ \bibinfo {author} {\bibfnamefont
  {J.}~\bibnamefont {\ifmmode~\check{Z}\else \v{Z}\fi{}elezn\'y}},\ }\bibfield
  {title} {\bibinfo {title} {Efficient electrical spin splitter based on
  nonrelativistic collinear antiferromagnetism},\ }\href
  {https://doi.org/10.1103/PhysRevLett.126.127701} {\bibfield  {journal}
  {\bibinfo  {journal} {Phys. Rev. Lett.}\ }\textbf {\bibinfo {volume} {126}},\
  \bibinfo {pages} {127701} (\bibinfo {year} {2021})}\BibitemShut {NoStop}%
\bibitem [{\citenamefont {Gonzalez~Betancourt}\ \emph
  {et~al.}(2023)\citenamefont {Gonzalez~Betancourt}, \citenamefont
  {Zub\'a\ifmmode~\check{c}\else \v{c}\fi{}}, \citenamefont
  {Gonzalez-Hernandez}, \citenamefont {Geishendorf}, \citenamefont {\ifmmode
  \check{S}\else \v{S}\fi{}ob\'a\ifmmode~\check{n}\else \v{n}\fi{}},
  \citenamefont {Springholz}, \citenamefont {Olejn\'{\i}k}, \citenamefont
  {\ifmmode~\check{S}\else \v{S}\fi{}mejkal}, \citenamefont {Sinova},
  \citenamefont {Jungwirth}, \citenamefont {Goennenwein}, \citenamefont
  {Thomas}, \citenamefont {Reichlov\'a}, \citenamefont {\ifmmode~\check{Z}\else
  \v{Z}\fi{}elezn\'y},\ and\ \citenamefont
  {Kriegner}}]{PhysRevLett.130.036702}%
  \BibitemOpen
  \bibfield  {author} {\bibinfo {author} {\bibfnamefont {R.~D.}\ \bibnamefont
  {Gonzalez~Betancourt}}, \bibinfo {author} {\bibfnamefont {J.}~\bibnamefont
  {Zub\'a\ifmmode~\check{c}\else \v{c}\fi{}}}, \bibinfo {author} {\bibfnamefont
  {R.}~\bibnamefont {Gonzalez-Hernandez}}, \bibinfo {author} {\bibfnamefont
  {K.}~\bibnamefont {Geishendorf}}, \bibinfo {author} {\bibfnamefont
  {Z.}~\bibnamefont {\ifmmode \check{S}\else
  \v{S}\fi{}ob\'a\ifmmode~\check{n}\else \v{n}\fi{}}}, \bibinfo {author}
  {\bibfnamefont {G.}~\bibnamefont {Springholz}}, \bibinfo {author}
  {\bibfnamefont {K.}~\bibnamefont {Olejn\'{\i}k}}, \bibinfo {author}
  {\bibfnamefont {L.}~\bibnamefont {\ifmmode~\check{S}\else \v{S}\fi{}mejkal}},
  \bibinfo {author} {\bibfnamefont {J.}~\bibnamefont {Sinova}}, \bibinfo
  {author} {\bibfnamefont {T.}~\bibnamefont {Jungwirth}}, \bibinfo {author}
  {\bibfnamefont {S.~T.~B.}\ \bibnamefont {Goennenwein}}, \bibinfo {author}
  {\bibfnamefont {A.}~\bibnamefont {Thomas}}, \bibinfo {author} {\bibfnamefont
  {H.}~\bibnamefont {Reichlov\'a}}, \bibinfo {author} {\bibfnamefont
  {J.}~\bibnamefont {\ifmmode~\check{Z}\else \v{Z}\fi{}elezn\'y}},\ and\
  \bibinfo {author} {\bibfnamefont {D.}~\bibnamefont {Kriegner}},\ }\bibfield
  {title} {\bibinfo {title} {Spontaneous anomalous hall effect arising from an
  unconventional compensated magnetic phase in a semiconductor},\ }\href
  {https://doi.org/10.1103/PhysRevLett.130.036702} {\bibfield  {journal}
  {\bibinfo  {journal} {Phys. Rev. Lett.}\ }\textbf {\bibinfo {volume} {130}},\
  \bibinfo {pages} {036702} (\bibinfo {year} {2023})}\BibitemShut {NoStop}%
\bibitem [{\citenamefont {Mazin}\ \emph {et~al.}(2023)\citenamefont {Mazin},
  \citenamefont {Gonz{\'a}lez-Hern{\'a}ndez},\ and\ \citenamefont
  {{\v{S}}mejkal}}]{mazin2023induced}%
  \BibitemOpen
  \bibfield  {author} {\bibinfo {author} {\bibfnamefont {I.}~\bibnamefont
  {Mazin}}, \bibinfo {author} {\bibfnamefont {R.}~\bibnamefont
  {Gonz{\'a}lez-Hern{\'a}ndez}},\ and\ \bibinfo {author} {\bibfnamefont
  {L.}~\bibnamefont {{\v{S}}mejkal}},\ }\bibfield  {title} {\bibinfo {title}
  {Induced monolayer altermagnetism in {MnP(S,Se)$_3$} and {FeSe}},\
  }\href@noop {} {\bibfield  {journal} {\bibinfo  {journal} {arXiv preprint
  arXiv:2309.02355}\ } (\bibinfo {year} {2023})}\BibitemShut {NoStop}%
\bibitem [{\citenamefont {Cui}\ \emph {et~al.}(2023)\citenamefont {Cui},
  \citenamefont {Zhu}, \citenamefont {Yao}, \citenamefont {Cui},\ and\
  \citenamefont {Yang}}]{PhysRevB.108.024410}%
  \BibitemOpen
  \bibfield  {author} {\bibinfo {author} {\bibfnamefont {Q.}~\bibnamefont
  {Cui}}, \bibinfo {author} {\bibfnamefont {Y.}~\bibnamefont {Zhu}}, \bibinfo
  {author} {\bibfnamefont {X.}~\bibnamefont {Yao}}, \bibinfo {author}
  {\bibfnamefont {P.}~\bibnamefont {Cui}},\ and\ \bibinfo {author}
  {\bibfnamefont {H.}~\bibnamefont {Yang}},\ }\bibfield  {title} {\bibinfo
  {title} {Giant spin-hall and tunneling magnetoresistance effects based on a
  two-dimensional nonrelativistic antiferromagnetic metal},\ }\href
  {https://doi.org/10.1103/PhysRevB.108.024410} {\bibfield  {journal} {\bibinfo
   {journal} {Phys. Rev. B}\ }\textbf {\bibinfo {volume} {108}},\ \bibinfo
  {pages} {024410} (\bibinfo {year} {2023})}\BibitemShut {NoStop}%
\bibitem [{\citenamefont {S{\o}dequist}\ and\ \citenamefont
  {Olsen}(2024)}]{sodequist2024two}%
  \BibitemOpen
  \bibfield  {author} {\bibinfo {author} {\bibfnamefont {J.}~\bibnamefont
  {S{\o}dequist}}\ and\ \bibinfo {author} {\bibfnamefont {T.}~\bibnamefont
  {Olsen}},\ }\bibfield  {title} {\bibinfo {title} {{Two-dimensional
  altermagnets from high throughput computational screening: Symmetry
  requirements, chiral magnons, and spin-orbit effects}},\ }\href@noop {}
  {\bibfield  {journal} {\bibinfo  {journal} {Applied Physics Letters}\
  }\textbf {\bibinfo {volume} {124}},\ \bibinfo {pages} {182409} (\bibinfo
  {year} {2024})}\BibitemShut {NoStop}%
\bibitem [{\citenamefont {Brekke}\ \emph {et~al.}(2023)\citenamefont {Brekke},
  \citenamefont {Brataas},\ and\ \citenamefont
  {Sudb\o{}}}]{PhysRevB.108.22442}%
  \BibitemOpen
  \bibfield  {author} {\bibinfo {author} {\bibfnamefont {B.}~\bibnamefont
  {Brekke}}, \bibinfo {author} {\bibfnamefont {A.}~\bibnamefont {Brataas}},\
  and\ \bibinfo {author} {\bibfnamefont {A.}~\bibnamefont {Sudb\o{}}},\
  }\bibfield  {title} {\bibinfo {title} {Two-dimensional altermagnets:
  Superconductivity in a minimal microscopic model},\ }\href
  {https://doi.org/10.1103/PhysRevB.108.224421} {\bibfield  {journal} {\bibinfo
   {journal} {Phys. Rev. B}\ }\textbf {\bibinfo {volume} {108}},\ \bibinfo
  {pages} {224421} (\bibinfo {year} {2023})}\BibitemShut {NoStop}%
\bibitem [{\citenamefont {Mermin}\ and\ \citenamefont
  {Wagner}(1966)}]{PhysRevLett.17.1133}%
  \BibitemOpen
  \bibfield  {author} {\bibinfo {author} {\bibfnamefont {N.~D.}\ \bibnamefont
  {Mermin}}\ and\ \bibinfo {author} {\bibfnamefont {H.}~\bibnamefont
  {Wagner}},\ }\bibfield  {title} {\bibinfo {title} {Absence of ferromagnetism
  or antiferromagnetism in one- or two-dimensional isotropic heisenberg
  models},\ }\href {https://doi.org/10.1103/PhysRevLett.17.1133} {\bibfield
  {journal} {\bibinfo  {journal} {Phys. Rev. Lett.}\ }\textbf {\bibinfo
  {volume} {17}},\ \bibinfo {pages} {1133} (\bibinfo {year}
  {1966})}\BibitemShut {NoStop}%
\bibitem [{\citenamefont {Milivojevi{\'c}}\ \emph {et~al.}(2024)\citenamefont
  {Milivojevi{\'c}}, \citenamefont {Orozovi{\'c}}, \citenamefont {Picozzi},
  \citenamefont {Gmitra},\ and\ \citenamefont
  {Stavri{\'c}}}]{milivojevic2024interplay}%
  \BibitemOpen
  \bibfield  {author} {\bibinfo {author} {\bibfnamefont {M.}~\bibnamefont
  {Milivojevi{\'c}}}, \bibinfo {author} {\bibfnamefont {M.}~\bibnamefont
  {Orozovi{\'c}}}, \bibinfo {author} {\bibfnamefont {S.}~\bibnamefont
  {Picozzi}}, \bibinfo {author} {\bibfnamefont {M.}~\bibnamefont {Gmitra}},\
  and\ \bibinfo {author} {\bibfnamefont {S.}~\bibnamefont {Stavri{\'c}}},\
  }\bibfield  {title} {\bibinfo {title} {Interplay of altermagnetism and weak
  ferromagnetism in two-dimensional {RuF$_4$}},\ }\href@noop {} {\bibfield
  {journal} {\bibinfo  {journal} {arXiv preprint arXiv:2401.15424}\ } (\bibinfo
  {year} {2024})}\BibitemShut {NoStop}%
\bibitem [{\citenamefont {Jain}\ \emph {et~al.}(2013)\citenamefont {Jain},
  \citenamefont {Ong}, \citenamefont {Hautier}, \citenamefont {Chen},
  \citenamefont {Richards}, \citenamefont {Dacek}, \citenamefont {Cholia},
  \citenamefont {Gunter}, \citenamefont {Skinner}, \citenamefont {Ceder},\ and\
  \citenamefont {Persson}}]{10.1063/1.4812323}%
  \BibitemOpen
  \bibfield  {author} {\bibinfo {author} {\bibfnamefont {A.}~\bibnamefont
  {Jain}}, \bibinfo {author} {\bibfnamefont {S.~P.}\ \bibnamefont {Ong}},
  \bibinfo {author} {\bibfnamefont {G.}~\bibnamefont {Hautier}}, \bibinfo
  {author} {\bibfnamefont {W.}~\bibnamefont {Chen}}, \bibinfo {author}
  {\bibfnamefont {W.~D.}\ \bibnamefont {Richards}}, \bibinfo {author}
  {\bibfnamefont {S.}~\bibnamefont {Dacek}}, \bibinfo {author} {\bibfnamefont
  {S.}~\bibnamefont {Cholia}}, \bibinfo {author} {\bibfnamefont
  {D.}~\bibnamefont {Gunter}}, \bibinfo {author} {\bibfnamefont
  {D.}~\bibnamefont {Skinner}}, \bibinfo {author} {\bibfnamefont
  {G.}~\bibnamefont {Ceder}},\ and\ \bibinfo {author} {\bibfnamefont {K.~a.}\
  \bibnamefont {Persson}},\ }\bibfield  {title} {\bibinfo {title} {{The
  Materials Project: A materials genome approach to accelerating materials
  innovation}},\ }\href {https://doi.org/10.1063/1.4812323} {\bibfield
  {journal} {\bibinfo  {journal} {APL Materials}\ }\textbf {\bibinfo {volume}
  {1}},\ \bibinfo {pages} {011002} (\bibinfo {year} {2013})}\BibitemShut
  {NoStop}%
\bibitem [{\citenamefont {Kopsk{\`y}}\ and\ \citenamefont
  {Litvin}(2002)}]{kopsky2002international}%
  \BibitemOpen
  \bibfield  {author} {\bibinfo {author} {\bibfnamefont {V.}~\bibnamefont
  {Kopsk{\`y}}}\ and\ \bibinfo {author} {\bibfnamefont {D.}~\bibnamefont
  {Litvin}},\ }\href@noop {} {\emph {\bibinfo {title} {International tables for
  crystallography, volume E: Subperiodic groups}}}\ (\bibinfo  {publisher}
  {Wiley Online Library},\ \bibinfo {year} {2002})\BibitemShut {NoStop}%
\bibitem [{\citenamefont {Chen}\ \emph
  {et~al.}(2023{\natexlab{b}})\citenamefont {Chen}, \citenamefont {Wang},
  \citenamefont {Li},\ and\ \citenamefont {Sanyal}}]{chen2023giant}%
  \BibitemOpen
  \bibfield  {author} {\bibinfo {author} {\bibfnamefont {X.}~\bibnamefont
  {Chen}}, \bibinfo {author} {\bibfnamefont {D.}~\bibnamefont {Wang}}, \bibinfo
  {author} {\bibfnamefont {L.}~\bibnamefont {Li}},\ and\ \bibinfo {author}
  {\bibfnamefont {B.}~\bibnamefont {Sanyal}},\ }\bibfield  {title} {\bibinfo
  {title} {Giant spin-splitting and tunable spin-momentum locked transport in
  room temperature collinear antiferromagnetic semimetallic cro monolayer},\
  }\href@noop {} {\bibfield  {journal} {\bibinfo  {journal} {Applied Physics
  Letters}\ }\textbf {\bibinfo {volume} {123}} (\bibinfo {year}
  {2023}{\natexlab{b}})}\BibitemShut {NoStop}%
\bibitem [{\citenamefont {Guo}\ \emph {et~al.}(2023)\citenamefont {Guo},
  \citenamefont {Liu},\ and\ \citenamefont {Lu}}]{guo2023quantum}%
  \BibitemOpen
  \bibfield  {author} {\bibinfo {author} {\bibfnamefont {P.-J.}\ \bibnamefont
  {Guo}}, \bibinfo {author} {\bibfnamefont {Z.-X.}\ \bibnamefont {Liu}},\ and\
  \bibinfo {author} {\bibfnamefont {Z.-Y.}\ \bibnamefont {Lu}},\ }\bibfield
  {title} {\bibinfo {title} {Quantum anomalous hall effect in collinear
  antiferromagnetism},\ }\href@noop {} {\bibfield  {journal} {\bibinfo
  {journal} {npj Computational Materials}\ }\textbf {\bibinfo {volume} {9}},\
  \bibinfo {pages} {70} (\bibinfo {year} {2023})}\BibitemShut {NoStop}%
\bibitem [{\citenamefont {Aroyo}(2016)}]{nespolo2017international}%
  \BibitemOpen
  \bibfield  {author} {\bibinfo {author} {\bibfnamefont {M.~I.}\ \bibnamefont
  {Aroyo}},\ }\href@noop {} {\emph {\bibinfo {title} {International Tables for
  Crystallography, Volume A: Space-group Symmetry}}}\ (\bibinfo  {publisher}
  {Wiley Online Library},\ \bibinfo {year} {2016})\BibitemShut {NoStop}%
\bibitem [{\citenamefont {Litvin}\ and\ \citenamefont
  {Wike}(2012)}]{litvin2012character}%
  \BibitemOpen
  \bibfield  {author} {\bibinfo {author} {\bibfnamefont {D.~B.}\ \bibnamefont
  {Litvin}}\ and\ \bibinfo {author} {\bibfnamefont {T.~R.}\ \bibnamefont
  {Wike}},\ }\href@noop {} {\emph {\bibinfo {title} {Character Tables and
  Compatibility Relations of the Eighty Layer Groups and Seventeen Plane
  Groups}}}\ (\bibinfo  {publisher} {Springer Science \& Business Media},\
  \bibinfo {year} {2012})\BibitemShut {NoStop}%
\bibitem [{\citenamefont {Kresse}\ and\ \citenamefont
  {Furthm{\"u}ller}(1996)}]{kresse1996efficiency}%
  \BibitemOpen
  \bibfield  {author} {\bibinfo {author} {\bibfnamefont {G.}~\bibnamefont
  {Kresse}}\ and\ \bibinfo {author} {\bibfnamefont {J.}~\bibnamefont
  {Furthm{\"u}ller}},\ }\bibfield  {title} {\bibinfo {title} {Efficiency of
  ab-initio total energy calculations for metals and semiconductors using a
  plane-wave basis set},\ }\href@noop {} {\bibfield  {journal} {\bibinfo
  {journal} {Computational materials science}\ }\textbf {\bibinfo {volume}
  {6}},\ \bibinfo {pages} {15} (\bibinfo {year} {1996})}\BibitemShut {NoStop}%
\bibitem [{\citenamefont {Kresse}\ and\ \citenamefont
  {Furthm\"uller}(1996)}]{kresse1996efficient}%
  \BibitemOpen
  \bibfield  {author} {\bibinfo {author} {\bibfnamefont {G.}~\bibnamefont
  {Kresse}}\ and\ \bibinfo {author} {\bibfnamefont {J.}~\bibnamefont
  {Furthm\"uller}},\ }\bibfield  {title} {\bibinfo {title} {Efficient iterative
  schemes for ab initio total-energy calculations using a plane-wave basis
  set},\ }\href {https://doi.org/10.1103/PhysRevB.54.11169} {\bibfield
  {journal} {\bibinfo  {journal} {Phys. Rev. B}\ }\textbf {\bibinfo {volume}
  {54}},\ \bibinfo {pages} {11169} (\bibinfo {year} {1996})}\BibitemShut
  {NoStop}%
\bibitem [{\citenamefont {Kresse}\ and\ \citenamefont
  {Joubert}(1999)}]{kresse1999ultrasoft}%
  \BibitemOpen
  \bibfield  {author} {\bibinfo {author} {\bibfnamefont {G.}~\bibnamefont
  {Kresse}}\ and\ \bibinfo {author} {\bibfnamefont {D.}~\bibnamefont
  {Joubert}},\ }\bibfield  {title} {\bibinfo {title} {From ultrasoft
  pseudopotentials to the projector augmented-wave method},\ }\href
  {https://doi.org/10.1103/PhysRevB.59.1758} {\bibfield  {journal} {\bibinfo
  {journal} {Phys. Rev. B}\ }\textbf {\bibinfo {volume} {59}},\ \bibinfo
  {pages} {1758} (\bibinfo {year} {1999})}\BibitemShut {NoStop}%
\bibitem [{\citenamefont {Perdew}\ \emph {et~al.}(1996)\citenamefont {Perdew},
  \citenamefont {Burke},\ and\ \citenamefont
  {Ernzerhof}}]{perdew1996generalized}%
  \BibitemOpen
  \bibfield  {author} {\bibinfo {author} {\bibfnamefont {J.~P.}\ \bibnamefont
  {Perdew}}, \bibinfo {author} {\bibfnamefont {K.}~\bibnamefont {Burke}},\ and\
  \bibinfo {author} {\bibfnamefont {M.}~\bibnamefont {Ernzerhof}},\ }\bibfield
  {title} {\bibinfo {title} {Generalized gradient approximation made simple},\
  }\href {https://doi.org/10.1103/PhysRevLett.77.3865} {\bibfield  {journal}
  {\bibinfo  {journal} {Phys. Rev. Lett.}\ }\textbf {\bibinfo {volume} {77}},\
  \bibinfo {pages} {3865} (\bibinfo {year} {1996})}\BibitemShut {NoStop}%
\bibitem [{\citenamefont {Dudarev}\ \emph {et~al.}(1998)\citenamefont
  {Dudarev}, \citenamefont {Botton}, \citenamefont {Savrasov}, \citenamefont
  {Humphreys},\ and\ \citenamefont {Sutton}}]{PhysRevB.57.1505}%
  \BibitemOpen
  \bibfield  {author} {\bibinfo {author} {\bibfnamefont {S.~L.}\ \bibnamefont
  {Dudarev}}, \bibinfo {author} {\bibfnamefont {G.~A.}\ \bibnamefont {Botton}},
  \bibinfo {author} {\bibfnamefont {S.~Y.}\ \bibnamefont {Savrasov}}, \bibinfo
  {author} {\bibfnamefont {C.~J.}\ \bibnamefont {Humphreys}},\ and\ \bibinfo
  {author} {\bibfnamefont {A.~P.}\ \bibnamefont {Sutton}},\ }\bibfield  {title}
  {\bibinfo {title} {Electron-energy-loss spectra and the structural stability
  of nickel oxide: An {LSDA+U} study},\ }\href
  {https://doi.org/10.1103/PhysRevB.57.1505} {\bibfield  {journal} {\bibinfo
  {journal} {Phys. Rev. B}\ }\textbf {\bibinfo {volume} {57}},\ \bibinfo
  {pages} {1505} (\bibinfo {year} {1998})}\BibitemShut {NoStop}%
\end{thebibliography}%

\end{document}